\documentclass[12pt]{iopart}

\newcommand{\mb}[1]{\mbox{\boldmath $#1$}}

\begin{document}

\jl{6}

\title{Equilibrium Configurations in the Dynamics of Irrotational
Dust Matter}

\author{Carlos F. Sopuerta} 

\address{Relativity and Cosmology Group, School of Computer Science 
and Mathematics, \\
University of Portsmouth, Portsmouth~PO1~2EG, England}

\address{E-mail: {\tt carlos.sopuerta@port.ac.uk}}

\begin{abstract}
Irrotational dust solutions of Einstein's equations are
suitable models to describe the general-relativistic aspects
of the gravitational instability mechanism for the formation of
cosmic structures.  In this paper we study their state space
by considering the local initial-value problem formulated in the
covariant fluid approach.  We consider a wide
range of models, from homogeneous and isotropic to highly
inhomogeneous irrotational dust models, showing how they
constitute equilibrium configurations (invariant sets) of the dynamics.  
Moreover, we give the characterization of such configurations, 
which provides an initial-data characterization of the models 
under consideration.
\end{abstract}





~

%
%

\section{Introduction}\label{sec1}
Current theories of the formation of cosmic structure explain structure
as the outcome of the amplification of small energy density fluctuations
left after the inflationary epoch. The usually proposed mechanism for
amplification is {\em gravitational instability}, which is essentially the
tendency of the self-gravity of a density fluctuation to induce 
collapse.  Within the framework of Newtonian Cosmology several 
successful techniques have been set up to describe this dynamical 
process (see~\cite{ASTR,SACO} and references therein).
Some remarkable and illustrative results are:
(i) The {\em Jeans instability}~\cite{JEAN}.  For small density
fluctuations there is a critical wavelength, the Jeans wavelength
$\lambda_J\,,$ such that fluctuations with a greater wavelength will
grow with time producing collapse, whereas fluctuations with a
smaller wavelength will be dispersed as sound waves.
(ii) The {\em Lin-Mestel-Shu instability}~\cite{LIMS}.
Given a self-gravitating, uniform, non-rotating, and pressure-free
distribution of matter at rest, spherically-symmetric collapse
takes place through a series of spherically-symmetric states.
However, the slightest departure from sphericity is systematically
magnified, giving rise to an instability.  In particular, if the
initial distribution is composed of oblate spheroids, the system
will approach an infinitely thin disk ({\em pancake} formation),
whereas if it is composed of prolate spheroids, the system will
approach an infinitely thin cylinder ({\em filament} formation).
Other similar studies can be found, e.g., in~\cite{ICKE,WHSI}.
(iii) The {\em Zel'dovich approximation}~\cite{ZELD}.  This
perturbative approach and its improvements (see, 
e.g.,~\cite{ASTR,SACO}) allow one to follow analytically the
growth of density fluctuations into the non-linear regime and
until caustics form.  It predicts that the generic outcome of
collapse is the formation of pancakes, although other structures 
like filaments or {\em clumps} (point-like singularities) can also 
form, but this is less probable.

However, despite the fact that Newtonian Cosmology provides a
wide-ranging picture of the dynamical process of the formation of
cosmic structures,  there are situations in which it cannot make
reliable predictions.  Of particular importance is the evolution of
superhorizon inhomogeneities, where the causal structure of the
Newtonian theory (in which the gravitational interaction is transmitted
instantaneously) is not suitable (see, e.g.,~\cite{SACO,SUHO}).
Another interesting issue for which the relativistic approach is needed
is the generation of gravitational waves by non-linear 
perturbations~\cite{MATA}.
Therefore, the relativistic study of the gravitational instability
mechanism is well motivated.  On the other hand, the relativistic theory
involves approaches that, from a technical point of view, are very different
from the Newtonian ones.  For instance, in order to use numerical methods
with a fluid approach we would need to consider an initial-value problem
formulation and to build a numerical code for the evolution equations.
In the Newtonian theory we would need to solve Poisson's equation at each
time whereas in General Relativity we only need to solve the constraints
initially and this ensures that they will be satisfied for all later times.

Most of the relativistic approaches make assumptions on the 
geometric structure of the spacetime, like the imposition of symmetries 
(e.g., spherical symmetry),  that make it impossible to obtain generic
results like those
described above within the Newtonian framework.
In this sense,  the covariant fluid approach~\cite{EHLE,COFA} is an
adequate starting point since it deals only with physical variables.
Several studies have appeared within this framework.  In~\cite{SILE},
Irrotational Dust Models (IDMs) were studied assuming only that
the magnetic part of the Weyl tensor, $H_{ab}\,,$ was negligible.
The dynamical analysis favoured the {\em spindle} as the generic outcome
of the collapse~\cite{BMPP}, in contrast to the Newtonian prediction, the
{\em pancake}.  However, the study of the dynamical consistency of the
constraints led to a conjecture that only some known models were
included~\cite{CONJ} [the Friedmann-Lema\^{\i}tre-Robertson-Walker (FLRW),
Bianchi I, and Szekeres dust models], leaving this question open.
In~\cite{SMEL} it was shown that if we relax the condition
$H_{ab}=0\,,$ assuming instead $H_{ab}$ to be transverse, $D^bH_{ab}=0$, the
whole set of constraints is, in general, unstable (a chain of infinite
new constraints appears).  The main conclusion we can extract is that the
dynamics of the gravitational instability mechanism is non-local
(see also~\cite{KOPO}), and hence we must be careful when imposing
conditions on $E_{ab}$ and $H_{ab}$~\cite{CONJ,SMEL,MLE1} (other studies
in this direction can be found in~\cite{IDMG}).

This suggests that the use of exact analytic approaches to understand the
gravitational instability mechanism is quite unrealistic. Therefore,
we should consider perturbative and/or numerical methods.
The aim of this paper is to contribute to the development of the latter.
To that end, we will consider IDMs.  Thus, we assume that pressure is
negligible and that the matter fluid flow is irrotational.  These assumptions
are analogous to those made in the original Zel'dovich
approximation~\cite{ZELD}.
Here, we will study equilibrium configurations of the dynamics
(IDMs constitute an infinite-dimensional dynamical system), as we will
call the invariant sets of the dynamics.
We will see that some exact IDMs, from homogeneous and isotropic to
inhomogeneous models, constitute equilibrium configurations and we will
find their characterization.   On one hand, this study provides a deeper
knowledge of the dynamics, especially of the role of the gradients in the
evolution equations,  responsible for the non-locality of the dynamics,
and which were neglected in most of the particular models considerer until
now in the literature.  On the other hand, the characterization of particular
IDMs as equilibrium configurations is at the same time an initial-data
characterization.  Then we know the initial data whose development
corresponds to those IDMs,  which provides interesting examples to check
numerical codes and also information about what kind of data we must not
prescribe if we want to study new behaviours. Moreover, it can be used
to identify attractors, repellers or asymptotic states in the dynamics.
Furthermore, an initial-data characterization provides also a solution of
the constraint equations, the solving of which presents in general strong
difficulties, especially if we are interested in prescribing generic initial
data.  This is even more important in cosmology, where we do not have
physically reasonable boundary conditions in order to get a well-posed
elliptic problem from the constraints.   In that respect, the information
that our study provides can be useful for solving the constraints in other
situations.

The plan of the paper is as follows: In Sec.~\ref{sec2}, using
the covariant fluid approach and tetrad methods, and
following~\cite{FOCO,VEEL}, we introduce a local Initial-Value Problem
(IVP) formulation for IDMs. In Sec.~\ref{sec3}, using this formulation,
we see that a wide range of exact IDMs constitute
equilibrium configurations of the dynamics.  Moreover, we find the
characterization of those states, which will provide an initial-data 
characterization of the irrotational dust models considered.  
We end with some comments and remarks in Sec.~\ref{sec5}.
The notation we use in this paper follows that of previous
works, especially~\cite{MAAR}.  We use units in which
$8\pi G=c=1$, round brackets enclosing indices denote
symmetrization and square brackets antisymmetrization.
Throughout the paper we use coordinate charts as well as
tetrads in order to express tensorial components.  The conventions
for indices are the following: we denote spacetime coordinate indices by
the lower-case Latin letters $a,\ldots,l=0,1,2,3$, and spacetime indices
with respect to an arbitrary basis 
$\{\mb{e}_0,\mb{e}_1,\mb{e}_2,\mb{e}_3\}$ by the remaining of lower-case
Latin letters $m,\ldots, z=0,1,2,3$. When we choose a basis adapted to
the dust velocity field, i.e., $\mb{e}_0=\mb{u}$, we will denote indices
with respect to a spatial basis $\{\mb{e}_1,\mb{e}_2,\mb{e}_3\}$ by
lower-case Greek letters $\alpha,\ldots, \lambda=1,2,3$ and spatial
coordinate indices by the remaining of lower-case Greek letters $\mu,\ldots,
\omega=1,2,3$.

\section{An initial-value problem description of irrotational dust
models}\label{sec2}

Irrotational dust solutions of Einstein's equations are models suitable
for describing the gravitational instability mechanism and the dynamics
of the Universe in the matter-dominated era.  Their energy-momentum
distribution is completely described by the fluid velocity $\mb{u}$ and
the energy density associated with it, $\rho$, the energy-momentum tensor
then being
\[ T_{ab} = \rho  u_au_b\,,~~~~ u^au_a=-1\,,~~~~ \rho > 0 \,.  \]
The fluid flow is irrotational ($u_{[a}\nabla_bu_{c]}=0$), i.e. $\mb{u}$
generates orthogonal spacelike hypersurfaces, and geodesic
($u^b\nabla_bu^a=0$) by virtue of the vanishing of the pressure
and the energy-momentum conservation equations.
Hence, there is locally a function $\tau(x^a)$ such the fluid velocity
is given by
\[ \mb{\vec{u}} = \frac{\mb{\partial}}{\mb{\partial\tau}}\,,
\hspace{10mm} \mb{u} = -\mb{d\tau} \,. \]
Then, $\tau$ is at the same time the proper time of the matter and the
label of the hypersurfaces orthogonal to $\mb{u}$,
$\Sigma(\tau_1)\equiv\{\tau(x^a)=\tau_1: \mbox{constant}\}\,.$
Choosing three independent first integrals of $\mb{u}$, $y^\mu(x^a)$
($u^a\partial_ay^\mu=0$), $\{\tau,y^\mu\}$ is a set of comoving geodesic
normal coordinates.  The line element in such a coordinate system has the
following form
\begin{equation}
ds^2 = -d\tau^2 + h_{\mu\nu}(\tau,y^\sigma)dy^\mu dy^\nu \,,
\label{liel}
\end{equation}
where $h_{\mu\nu}(\tau,y^\sigma)$ are the non-zero components of
the orthogonal projector to the fluid velocity, $h_{ab}\equiv g_{ab}+u_au_b$,
whose restriction to a hypersurface $\Sigma(\tau)$ coincides with its
first fundamental form.

In order to establish an IVP for IDMs in terms of variables with a clear
physical and geometrical meaning, it will be very convenient to describe
them in terms of the
covariant fluid approach introduced by Ehlers~\cite{EHLE} (see~\cite{COFA}
for more details) and using an orthonormal basis adapted to the fluid
velocity, $\{\mb{e}_0=\mb{u},\mb{e}_1,\mb{e}_2,\mb{e}_3\}$
[$\mb{e}_m\cdot\mb{e}_n=\eta_{mn}\equiv\mbox{diag} (-1,1,1,1)$].
We can partially fix the freedom in the choice of the triad
$\{\mb{e}_\alpha\}$ by requiring it to be parallely propagated along
the fluid flow
\[ \dot{\mbox{e}}_\alpha{}^a = u^b\nabla_b\mbox{e}_\alpha{}^a =0\,. \]
This choice makes the local angular velocity vanish
\[ \Omega^\alpha\equiv\textstyle{1\over2}\varepsilon^{\alpha\beta
\delta}\mb{e}_\beta\cdot\dot{\mb{e}}_\delta = 0\,. \]
The advantage of this choice is that we avoid having as a variable
$\Omega^\alpha$, which is not a dynamical quantity since there is no
evolution equation for it.

The variables that we will use to describe IDMs can be divided into
five groups (see~\cite{MAVU,WAEL} and references therein for details):
(i) {\it Metric variables}. The components of the triad vectors
in adapted coordinates $\{\tau,y^\mu\}$ [see Eq.~(\ref{liel})],
$\mbox{e}_\alpha{}^\mu\,.$  (ii) {\it Connection variables}.
The spatial commutators, $\gamma^\alpha{}_{\beta\lambda}$, defined by
the commutation relations between the triad vectors,
\begin{equation}
\left[\vec{\mbox{e}}_\beta,\vec{\mbox{e}}_\lambda\right]=
\gamma^\alpha{}_{\beta\lambda}\vec{\mbox{e}}_\alpha\,,\hspace{5mm}
\gamma^\alpha{}_{[\beta\lambda]}=\gamma^\alpha{}_{\beta\lambda}\,.
\label{spco}
\end{equation}
In this work we will also use the equivalent variables introduced by
Sch\"ucking, Kundt and Behr (see~\cite{ELMC} and references therein)
\[  \gamma^\alpha{}_{\beta\lambda} = 2 a_{[\beta}\delta^\alpha{}_{\lambda]}
+\varepsilon_{\beta\lambda\delta}n^{\alpha\delta}~~
\Longleftrightarrow ~~
 a_\alpha=\textstyle{1\over2}\gamma^\beta{}_{\alpha\beta} \,,
\hspace{4mm} n^{\alpha\beta}=\textstyle{1\over2}
\varepsilon^{\lambda\delta(\alpha}\gamma^{\beta)}{}_{\lambda\delta}\,.\]
(iii) {\it Kinematical variables}. The expansion $\Theta$
($\equiv\nabla_au^a$) and the shear tensor of the fluid worldlines.
The shear is a symmetric and trace-free {\em spatial}~\cite{SPAT} tensor
\[ \sigma_{ab}\equiv h_{(a}{}^ch_{b)}{}^d\nabla_du_c-(\Theta/3)h_{ab}
= \nabla_au_b-(\Theta/3)h_{ab} \,. \]
It is worth noting that the restriction of the quantity
\begin{equation}
K_{ab}\equiv (\Theta/3)h_{ab}+\sigma_{ab} \,, \label{sffs}
\end{equation}
to a hypersurface $\Sigma(\tau)$ coincides with its second fundamental
form.  In this work we will consider as variables the expansion, $\Theta\,$ 
and the five independent components of the shear in a triad,
$\sigma_{\alpha\beta}\,,$ or their following representation:
\begin{equation}
\fl \sigma_+\equiv -\textstyle{3\over2}\sigma_{11}\,, \hspace{3mm}
\sigma_-\equiv \textstyle{\sqrt{3}\over2}(\sigma_{22}-\sigma_{33})\,,
\hspace{3mm} \sigma_1\equiv\sqrt{3}\,\sigma_{23}\,, \hspace{3mm}
\sigma_2\equiv\sqrt{3}\,\sigma_{13}\,, \hspace{3mm}
\sigma_3\equiv\sqrt{3}\,\sigma_{12}\,. \label{irva}
\end{equation}
(iv) {\em Matter variables}. Apart from the fluid velocity,
the only matter variable in our case is the energy density $\rho$
($= T_{00}$).  The Ricci tensor is then determined through
Einstein's equations
\begin{equation}
R_{ab}-\textstyle{1\over2}R g_{ab}+\Lambda g_{ab}= \rho\,u_au_b \,,
\label{eins}
\end{equation}
where $\Lambda$ denotes the cosmological constant.
(v) {\em Weyl
tensor variables}.  The Weyl tensor $C_{abcd}$ describes
the spacetime curvature not determined locally by matter.
Its ten independent components can be divided into two spatial,
symmetric and trace-free tensors
\[ E_{ab}=C_{acbd}u^cu^d\,,\hspace{4mm}
H_{ab}=*C_{acbd}u^cu^d\hspace{2mm} (*C_{abcd}\equiv
\textstyle{1\over2}\eta_{ab}{}^{ef}C_{cdef}) \,, \]
where $\eta_{abcd}$ is the spacetime volume 4-form.
Using the analogy with electromagnetism, they are called the
gravito-electric and -magnetic fields respectively. Whereas the
gravito-electric field produces tidal forces having a Newtonian
analogue (the trace-free part of the Hessian of the Newtonian
potential), the gravito-magnetic field has no Newtonian
analogue.  The five independent components of the gravito-electric
and -magnetic fields can be represented by the quantities
($E_+$, $E_-$, $E_1$, $E_2$, $E_3$) and
($H_+$, $H_-$, $H_1$, $H_2$, $H_3$), which are defined as in~(\ref{irva}).

The equations governing the behaviour of these quantities come
from the Ricci identities applied to $\mb{u}$, the second Bianchi 
identities, Einstein's equations~(\ref{eins}) and the Gauss equations
applied to the foliation $\{\Sigma(\tau)\}$ 
(see~\cite{COFA} for details).  In the covariant fluid
approach the covariant derivative $\nabla_a$ is decomposed into a 
time derivative along the fluid velocity,  $\dot{A}^{a\cdots}{}_{b\cdots}
\equiv u^c\nabla_c A^{a\cdots}{}_{b\cdots}$, 
and a spatial covariant derivative tangent to the hypersurfaces 
$\Sigma(\tau)$, $D_c A^{a\cdots}{}_{b\cdots}\equiv 
h^a{}_e\cdots h^f{}_bh_c{}^d\nabla_d A^{e\cdots}{}_{f\cdots}$, 
where $A^{a\cdots}{}_{b\cdots}$ is an arbitrary tensor field.  It will be
useful to introduce the spatial divergence and curl of an arbitrary
2-index tensor $A_{ab}$~\cite{ANAL} (see~\cite{MAAR} for details on this
notation)
\[ \mbox{div}\,(A)_a\equiv D^bA_{ab}\,,\hspace{5mm}
\mbox{curl}(A)_{ab}\equiv\varepsilon_{cd(a}D^cA_{b)}{}^d \,, \]
where $\varepsilon_{abc}\equiv\eta_{abcd}u^d$ ($\varepsilon^{abc}
\varepsilon_{def}=3! h^{[a}{}_dh^b{}_eh^{c]}{}_f$) is the volume
3-form of the hypersurfaces $\Sigma(\tau)$.  The projection of
these definitions onto a triad $\{\mb{e}_\alpha\}$ gives\footnote{Angled
brackets on indices denote the spatially projected, symmetric and
tracefree part: $A_{\langle\alpha\beta\rangle}=A_{(\alpha\beta)}-
(A^\lambda{}_\lambda/3)\delta_{\alpha\beta}\,.$}
\[ \mbox{curl}(A)_{\alpha\beta}= \varepsilon^{\lambda\delta}
{}_{<\alpha}(\mb{\partial}_{|\lambda|}-a_{|\lambda|})A_{\beta>\delta}
+\textstyle{1\over2}n^\delta{}_\delta A_{\alpha\beta}-
3n_{<\alpha}{}^\delta A_{\beta>\delta} \,, \]
\[ \mbox{div}(A)_\alpha = (\mb{\partial}_\delta-3a_\delta)A^\delta
{}_\alpha-\varepsilon_\alpha{}^{\beta\delta}n_\beta{}^\lambda
A_{\lambda\delta} \,, \]
where $\mb{\partial}_\alpha\equiv\mbox{e}_\alpha{}^\mu\partial_{y^\mu}\,.$
Finally, we define the commutator of two spatial symmetric tensors,
$A_{ab}$ and $B_{ab}$, as
\[ [A,B]_{ab}\equiv 2A_{[a}{}^cB_{b]c} \,, \hspace{4mm} [A,B]_a\equiv
\textstyle{1\over2}\varepsilon_{abc}[A,B]^{bc}=\varepsilon_{abc}A^b{}_d
B^{cd} \,. \]

The set of dynamical equations can be divided into two groups:
(i) Evolution equations, which give the rate of change of our
variables along the fluid world-lines.
(ii) Constraint equations, which are relations only between spatial
derivatives of our variables.  The form of these equations, written in
an orthonormal basis adapted to the fluid velocity, is the following

~

\noindent{\em Evolution equations}

\begin{equation}
\dot{\mbox{e}}_\alpha{}^\mu = -\left(\textstyle{1\over3}\Theta
\delta_\alpha{}^\beta+\sigma_\alpha{}^\beta\right)
\mbox{e}_\beta{}^\mu\,, \label{base}
\end{equation}
\begin{equation}
\dot{\gamma}^\delta_{\alpha\beta} = \left(\textstyle{1\over3}\Theta
\delta^\epsilon{}_{[\alpha}+\sigma^\epsilon{}_{[\alpha}
\right)\gamma^\delta_{\beta]\epsilon}+
2\delta^{\delta\epsilon}\delta_{\lambda\gamma}
\delta^{(\kappa}{}_{[\alpha}\sigma^{\lambda)}{}_{\beta]}
\gamma^{\gamma}_{\epsilon\kappa}
-\varepsilon_{\alpha\beta\epsilon}H^{\delta\epsilon}\,, \label{comm}
\end{equation}
\begin{equation}
\dot{\Theta}= -\textstyle{1\over3}\Theta^2-\sigma^{\alpha\beta}
\sigma_{\alpha\beta}-\textstyle{1\over2}\rho+\Lambda \,, \label{expa}
\end{equation}
\begin{equation}
\dot{\sigma}_{\alpha\beta}=-\textstyle{2\over3}\Theta\sigma_{\alpha\beta}
-\sigma_{<\alpha}{}^\delta\sigma_{\beta>\delta}-E_{\alpha\beta} \,,
\label{shea}
\end{equation}
\begin{equation}
\dot{\rho}=-\Theta\rho \,, \label{mate}
\end{equation}
\begin{equation}
\dot{E}_{\alpha\beta}-\mbox{curl}(H)_{\alpha\beta} =
-\Theta E_{\alpha\beta}+3\sigma_{<\alpha}{}^\delta E_{\beta>\delta}-
\textstyle{1\over2}\rho\sigma_{\alpha\beta}\,, \label{ewey}
\end{equation}
\begin{equation}
\dot{H}_{\alpha\beta}+\mbox{curl}(E)_{\alpha\beta} =
-\Theta H_{\alpha\beta}+3\sigma_{<\alpha}{}^\delta H_{\beta>\delta}
\,,  \label{hwey}
\end{equation}

~

\noindent{\em Constraint equations}

\begin{equation}
{\cal C}^0{}_{\alpha\beta}{}^\mu\equiv\gamma^\delta_{\alpha\beta}
\mb{\partial}_\delta y^\mu -
[\mb{\partial}_\alpha,\mb{\partial}_\beta]y^\mu = 0 \,, \label{li1}
\end{equation}
\begin{equation}
{\cal C}^1{}_\alpha\equiv\mbox{div}(\sigma)_\alpha-
\textstyle{2\over3}\mb{\partial}_\alpha\Theta = 0 \,, \label{li3}
\end{equation}
\begin{equation}
{\cal C}^2{}_{\alpha\beta}\equiv\mbox{curl}(\sigma)_{\alpha\beta}-
H_{\alpha\beta} = 0 \,, \label{li4}
\end{equation}
\begin{equation}
{\cal C}^3{}_\alpha\equiv\mbox{div}(E)_\alpha-
\varepsilon_{\alpha\beta\delta}\sigma^{\beta\epsilon}
H^\delta{}_\epsilon-\textstyle{1\over3}\mb{\partial}_\alpha\rho = 0
\,, \label{li5}
\end{equation}
\begin{equation}
{\cal C}^4{}_\alpha\equiv\mbox{div}(H)_\alpha+
\varepsilon_{\alpha\beta\delta}\sigma^{\beta\epsilon}
E^\delta{}_\epsilon = 0 \,, \label{li6}
\end{equation}
\begin{equation}
{\cal C}^5\equiv \rho-\textstyle{1\over3}\Theta^2+\textstyle{1\over2}
\sigma^{\alpha\beta}\sigma_{\alpha\beta}-\textstyle{1\over2}
{}^3 R + \Lambda = 0 \,, \label{li7}
\end{equation}
\begin{equation}
{\cal C}^6{}_{\alpha\beta}\equiv E_{\alpha\beta}-{}^3S_{\alpha\beta}+
\sigma_{<\alpha}{}^\delta\sigma_{\beta>\delta}-\textstyle{1\over3}
\Theta\sigma_{\alpha\beta}= 0 \,. \label{li8}
\end{equation}
In these equations, ${}^3R$ and ${}^3S_{\alpha\beta}$ are
the scalar curvature and the trace-free part of the Ricci tensor of the
hypersurfaces $\Sigma(\tau)$ respectively.  They must be understood as
given in terms of $\gamma^\delta_{\alpha\beta}$ and their derivatives,
through the expression
\[ {}^3R_{\alpha\beta}=
\mb{\partial}_\lambda(\Gamma^\lambda_{\alpha\beta})-
\mb{\partial}_\beta(\Gamma^\lambda_{\alpha\lambda})+
\Gamma^\epsilon_{\alpha\beta}\Gamma^\lambda_{\epsilon\lambda}-
\Gamma^\epsilon_{\alpha\lambda}\Gamma^\lambda_{\epsilon\beta}+
\Gamma^\lambda_{\alpha\epsilon}\gamma^\epsilon_{\beta\lambda}\,, \]
where $\Gamma^\alpha_{\beta\delta}\equiv\mb{e}^\alpha\cdot
(\nabla^{}_{\mb{e}_\delta}\mb{e}_\beta)$ are the Ricci rotation
coefficients associated with the triad $\{\mb{e}_\alpha\}$, 
related to $\gamma^\alpha_{\beta\delta}$ by
$\delta_{\alpha\epsilon}\Gamma^\epsilon_{\beta\lambda} =
\delta_{\alpha\epsilon}\gamma^\epsilon_{\lambda\beta}+
\delta_{\beta\epsilon}\gamma^\epsilon_{\alpha\lambda}+
\delta_{\lambda\epsilon}\gamma^\epsilon_{\alpha\beta}\,.$

Now, we will study the evolution equations~(\ref{base}-\ref{hwey})
to show that they form a first-order symmetric hyperbolic system of
partial differential equations (general accounts of this subject can
be found in~\cite{BPDE}, and specializations in general relativity
related with this work in~\cite{SIGR,FRRE}).
Following~\cite{VEEL} we can write the evolution
equations~(\ref{base}-\ref{hwey}) in matrix form as
\begin{equation} 
{\cal M}^a \mb{\partial}_a \mb{U} = \mb{N} \,, \label{fosh}
\end{equation}
where we have grouped the variables (38 in all) into the vector $\mb{U}$ 
\begin{equation}
\mb{U} = (\mb{U}_{basis},\mb{U}_{connection},\mb{U}_{kinematical},
\mb{U}_{matter},\mb{U}_{Weyl})^T \,, \label{vecv}
\end{equation}
where the superscript ${}^T$ stands for the transpose of a row vector,
and where
\[ \mb{U}_{basis}=(e_1{}^\mu,e_2{}^\mu,e_3{}^\mu)^T\,, \] 
\[ \mb{U}_{connection}=(a_1,a_2,a_3,n_{11},n_{12},n_{13},n_{22},n_{23},
n_{33})^T\,, \]
\[ \mb{U}_{kinematical}=(\Theta,\sigma_+,\sigma_-,\sigma_1,\sigma_2,
\sigma_3)^T\,, \hspace{4mm} \mb{U}_{matter}=(\rho)\,, \]
\[ \mb{U}_{Weyl}=(E_+,E_-,E_1,E_2,E_3,H_+,H_-,H_1,H_2,H_3)^T \,. \]
The right-hand side of (\ref{fosh}) is the {\em principal part}
of the system of equations, whose form is determined by the matrices
${\cal M}^a$, which in our case are (see also~\cite{VEEL})
\[ {\cal M}^0 = \left( \begin{array}{ccccc}
\mbox{Id}_{9} & \cdot & \cdot & \cdot & \cdot  \\
\cdot & \mbox{Id}_9 & \cdot & \cdot & \cdot  \\
\cdot & \cdot & \mbox{Id}_6 & \cdot & \cdot  \\
\cdot & \cdot & \cdot & 1 & \cdot  \\
\cdot & \cdot & \cdot & \cdot & \mbox{Id}_{10} \end{array} \right)
\,, \hspace{4mm} 
{\cal M}^\alpha = \left( \begin{array}{ccccc}
\mbox{0}_{9} & \cdot & \cdot & \cdot & \cdot  \\
\cdot & \mbox{0}_9 & \cdot & \cdot & \cdot  \\
\cdot & \cdot & \mbox{0}_6 & \cdot & \cdot  \\
\cdot & \cdot & \cdot & 0 & \cdot  \\
\cdot & \cdot & \cdot & \cdot & \mbox{W}^\alpha \end{array} \right)
\,,\]
where $\mbox{Id}_N$ and $\mbox{0}_N$ denote the $N\times N$ identity
and zero matrices respectively, and where the matrices $\mbox{W}^\alpha$
are given by
\[ \mbox{W}^1 = \left( \begin{array}{cc}
\mbox{0}_5 & {\cal A} \\ 
-{\cal A} & \mbox{0}_5 \end{array} \right) \,, \hspace{4mm} 
\mbox{W}^2 = \left( \begin{array}{cc}
\mbox{0}_5 & {\cal B} \\ 
-{\cal B} & \mbox{0}_5 \end{array} \right) \,, \hspace{4mm} 
\mbox{W}^3 = \left( \begin{array}{cc}
\mbox{0}_5 & {\cal C} \\ 
-{\cal C} & \mbox{0}_5 \end{array} \right) \,, \hspace{4mm} \]
where
\[ \fl {\cal A} = \left( \begin{array}{ccccc}
\cdot & \cdot & \cdot & \cdot & \cdot  \\
\cdot & \cdot & \mbox{\small 1} & \cdot & \cdot  \\
\cdot & \mbox{\small -1} & \cdot & \cdot & \cdot  \\
\cdot & \cdot & \cdot & \cdot & \textstyle{-1\over2} \\
\cdot & \cdot & \cdot & \textstyle{1\over2} & \cdot \end{array} \right)
\,
{\cal B} = \left( \begin{array}{ccccc}
\cdot & \cdot & \cdot & \textstyle{\sqrt{3}\over2} & \cdot  \\
\cdot & \cdot & \cdot & \textstyle{-1\over2} & \cdot  \\
\cdot & \cdot & \cdot & \cdot & \textstyle{1\over2} \\
\textstyle{-\sqrt{3}\over2} & \textstyle{1\over2} & \cdot & \cdot & \cdot \\
\cdot & \cdot & \textstyle{-1\over2} & \cdot & \cdot  \end{array} \right)
\,
{\cal C} = \left( \begin{array}{ccccc}
\cdot & \cdot & \cdot & \cdot & \textstyle{-\sqrt{3}\over2} \\
\cdot & \cdot & \cdot & \cdot & \textstyle{-1\over2} \\
\cdot & \cdot & \cdot & \textstyle{-1\over2} & \cdot  \\
\cdot & \cdot & \textstyle{1\over2} & \cdot & \cdot  \\
\textstyle{\sqrt{3}\over2} & \textstyle{1\over2} & \cdot & \cdot & \cdot
\end{array}\right) . \]
As we can see, ${\cal A}$, ${\cal B}$, and ${\cal C}$ are antisymmetric,
and hence the matrices $\mb{W}^\alpha$ are symmetric.  Moreover, \mb{N}
is a vector which depends analytically on the variables $\mb{U}$.  

This formulation shows that the evolution system~(\ref{fosh}) satisfies
the conditions to be a first-order symmetric hyperbolic system (FOSHS) 
of partial differential equations (see~\cite{FOCO,VEEL}).  Indeed,
the matrices ${\cal M}^a$ are symmetric, and there is a 1-form $\mb{t}$ 
such that $t_a{\cal M}^a$ is a positive definite matrix; this is
immediate upon taking $\mb{t}=-\mb{u}$ (for more details see~\cite{BPDE}).
Moreover, if every 1-form $\mb{t}$ satisfying
the last property is past-directed, the system is said to be
{\em causal}.  We can check that (\ref{fosh}) is causal (see
also~\cite{VEEL}).  Furthermore, as we have said before, FOSHSs  
admit a well-posed IVP~\cite{BPDE}, which implies that given smooth 
initial data $\mb{{}_oU}\equiv\mb{U}|_{\Sigma(\tau_o)}\,,$
there exists a unique smooth solution $\mb{{}_\ast U}$.

Now let us consider the constraints.  First of all, we have to point out
that not all of them are independent.  We find the following relationships
between them
\[ {\cal C}^3{}_\alpha = \textstyle{1\over3}\Theta{\cal C}^1{}_\alpha-
\textstyle{1\over2}\sigma_{\alpha\beta}{\cal C}^1{}^\beta+
[\sigma,{\cal C}^2{}]_\alpha+\mbox{div}({\cal C}^6{})_\alpha-
\textstyle{1\over3}D_\alpha{\cal C}^5\,,  \]
\[ {\cal C}^4{}_\alpha = \textstyle{1\over2}\mbox{curl}({\cal C}^1){}_\alpha
-\mbox{div}({\cal C}^2{})_\alpha \,. \]
The last one was already given in~\cite{MAAR,VELD}.  This means that
if constraints ${\cal C}^1{}_\alpha$, ${\cal C}^2{}_{\alpha\beta}$,
${\cal C}^5$, and ${\cal C}^6{}_{\alpha\beta}$ are satisfied, constraints
${\cal C}^3{}_\alpha$ and ${\cal C}^4{}_\alpha$ will be satisfied
automatically.   Another important point concerning the constraints
is their consistency with evolution.  In ~\cite{MAAR,VELD}, consistency
was proved for analytical initial data.   In~\cite{FRRE}, a general
treatment for perfect fluids is given and consistency has also been shown
for smooth initial data, but the constraints have not been split
according to the variables used.  In our treatment, we can construct
a closed system of evolution equations for
$\mb{V}=(\mb{U},\mb{C}\equiv({\cal C}^A))$.  It is worth noting that
spatial derivatives only appear in the equations for ${\cal C}^3{}_\alpha$
and ${\cal C}^4{}_\alpha$, and they have essentially the same principal
part as Maxwell's equations.  Therefore,  the evolution equations found
for $\mb{V}$ form a FOSHS.  Then, given smooth initial data
$\mb{{}_oU}$ satisfying the constraints (${\cal C}^A[\mb{{}_oU}]=0$),
and with $\mb{{}_\ast U}$ being the only solution of~(\ref{fosh}), it is
clear that $(\mb{{}_\ast U},\mb{0})$ is a solution, and is unique.

\section{Equilibrium configurations and their characterization}
\label{sec3}

We have just seen that the dynamical equations for IDMs constitute
a symmetric hyperbolic system, which means we can formulate
a well-posed IVP for irrotational dust matter.  As a dynamical
system, one important question is the study of equilibrium
configurations in the state space, or in a more technical
terminology (see, e.g.,~\cite{DYSY}),  the study of invariant
sets in the evolution.  Here we will look for such
configurations, associated with particular IDMs, finding the relations
that characterize them, which also characterize the associated IDMs
in terms of initial data.

Let us start by introducing some ideas, concepts and the procedure that we
will use systematically.   First, the state of the system will be described
by the dynamical variables contained in the vector $\mb{U}$
[Eq.~(\ref{vecv})], which we will call the {\em state} vector.
Then, we will say that the state of the system at a time $\tau_1$ is
determined by $\mb{U}$ on $\Sigma(\tau_1)$, i.e.~$\mb{U}|_{\Sigma(\tau_1)}$.
On the other hand, by {\em configuration} we will mean sets of states
characterized by relations which do not involve explicitly time or time
derivatives. That is, relations of the form
\begin{equation}
{\cal R}^A[\mb{U},\mb{\partial}_\alpha\mb{U},
\mb{\partial}_\alpha\mb{\partial}_\beta\mb{U},\ldots] = 0\,. \label{conf}
\end{equation}
For instance, the constraints [Equations~(\ref{li1}-\ref{li8})] have this
form.   Then a particular configuration whose characteristic
relations~(\ref{conf}) are preserved by the evolution will be called
an {\em equilibrium configuration} (an invariant set). For these,
${\cal R}^A|_{\Sigma(\tau_1)}=0$ implies that ${\cal R}^A=0$ for
$\tau > \tau_1$. The constraints are again a good example.  A consistent set 
of relations of the form~(\ref{conf}), and defined in an open domain of the
spacetime, determines a particular IDM in this domain.  Therefore, this set
of relations constitutes both a spacetime and an initial-data
characterization of that IDM.

An interesting example is the configuration defined by the vanishing of 
the gravito-electric and -magnetic fields:
\[ {\cal R}^1{}_{ab}\equiv E_{ab}=0\,, \hspace{5mm}
{\cal R}^2{}_{ab}\equiv H_{ab}=0\,. \]
As is well-known, the vanishing of these two tensors in an open domain
of the spacetime implies that the metric there is a FLRW metric.
However, assuming that these quantities vanish only on an open
domain of a hypersurface $\Sigma(\tau_1)$,
$E_{ab}|_{\Sigma(\tau_1)}=H_{ab}|_{\Sigma(\tau_1)}=0$, this does not 
imply the vanishing of these quantities at a different time.  This
is a consequence of the evolution equations (\ref{ewey},\ref{hwey}): from
(\ref{ewey}) we see that when $\rho\neq 0$ the shear acts a source
generating gravito-electric field $E_{ab}$, and from (\ref{hwey}) we
see that $\mbox{curl}(E)_{ab}$ acts as a source for the gravito-magnetic 
field $H_{ab}$.  Therefore, we conclude that $E_{ab}=H_{ab}=0$, which 
is a spacetime characterization of the FLRW models, is not an 
equilibrium configuration.  This reflects the obvious fact that
a spacetime characterization is not in general an initial-data
characterization.  On the other hand, the vacuum configuration, defined by
\[ {\cal R}\equiv \rho = 0 \,, \]
is an equilibrium configuration as we can see from the evolution
equation~(\ref{mate}).   

To study whether a particular IDM determines an equilibrium
configuration we will look for a set of relations~(\ref{conf}),
$\mb{R}\equiv({\cal R}^A)$, satisfied by that IDM, and such that
they are preserved by the evolution. To that end, we will consider
the relations $\mb{R}$ as new variables and will study their evolution.
For all the particular IDMs we are going to consider in this paper, we
will look for evolution equations having the following form
\begin{equation}
\dot{\mb{R}}+{\cal O}^\alpha[\mb{U},\mb{R}]\mb{\partial}_\alpha\mb{U}
+ {\cal P}^\alpha[\mb{U},\mb{R}]\mb{\partial}_\alpha\mb{R} =
\mb{X}[\mb{U},\mb{R}]  \,. \label{sysc}
\end{equation}
and with the property
\[ {\cal O}^\alpha[\mb{U},\mb{0}]=0\,, \hspace{3mm}
\mb{X}[\mb{U},\mb{0}]=\mb{0}\,. \]
As is clear, this property implies that $\mb{R}=\mb{0}$ is a solution.
Then, if we can show that this is the only possible solution, it follows
that the relations in $\mb{R}$ determine an equilibrium configuration.
One way to tackle this question
is to consider the variables $\mb{V}\equiv(\mb{U},\mb{R})$ and their
evolution equations~(\ref{fosh},\ref{sysc}).   In all the cases we will
consider here, the coefficients appearing in these equations
are analytic functions of the variables $\mb{V}$.  Therefore, the
system~(\ref{fosh},\ref{sysc}) for $\mb{V}$ will satisfy the conditions
of the Cauchy-Kowaleski theorem, which does not assume hyperbolicity
at all (a statement of this theorem can be found, e.g., 
in~\cite{BPDE}), and hence, the uniqueness of
the solution $\mb{R}=\mb{0}$ is ensured for analytical initial data
(assuming, of course, this initial data $\mb{{}_oU}$ is such that
$\mb{{}_oR}=\mb{0}$).  In the cases where the complete system
(\ref{fosh},\ref{sysc}) is symmetric or strongly~\cite{STRH} hyperbolic
(see, e.g.,~\cite{BPDE}) we will get a characterization also
valid for smooth initial data.  The constraints constitute again an  
example since we obtain a FOSHS for them.

In what follows we will apply this procedure to the following IDMs:
In~Sec.~\ref{flrw}, to the FLRW models. In~Sec.~\ref{spho} to the spatially
homogeneous IDMs and the Bianchi I subcase.  In~Sec.~\ref{szmo} to the
Szekeres models.  Finally, in~Sec.~\ref{plam}, to the IDMs in which the
hypersurfaces $\Sigma(\tau)$ are flat.  In all these cases we will
give the dynamical system for $\mb{V}$ that shows explicitly that 
$\mb{R}=\mb{0}$ is a solution of the evolution equations for the 
relations~(\ref{conf}) which determine the equilibrium configuration.
However, in some cases this system for the variables $\mb{V}$ will not 
provide a well-posed IVP for smooth initial data.  In these cases we can
use the fact that the relations $\mb{R}$ are defined in terms of
$\mb{U}$, and hence we can get several different systems of evolution
equations for $\mb{R}$.  Moreover, we can also use the fact that 
equations~(\ref{fosh}) are not coupled with the equations for $\mb{R}$, 
and that given smooth initial data $\mb{{}_oU}\,,$ there exists a unique
smooth solution ${}_\ast\mb{U}(\tau,y^\alpha)$ such that
${}_\ast\mb{U}|_{\Sigma(\tau_o)}=\mb{{}_oU}$.  Then given a system
of equations for $\mb{R}\,,$ we can substitute $\mb{U}$ for ${}_\ast\mb{U}$
in them so that we get a linear system of partial differential
equations
\[ \dot{\mb{R}}+ {\cal Q}^\mu[\tau,y^\nu]\partial_\mu\mb{R} =
\mb{Y}[\tau,y^\nu,\mb{R}]\,, \]
where all the coefficients are smooth.  If we can obtain a
system like this such that it admits the formulation of a well-posed
IVP for smooth initial data, and if we can show that $\mb{R}=0$ is a
solution, then we would have shown that it is the solution.  Therefore,
our characterization would be also valid for smooth initial data.
We have applied this procedure successfully to some of the cases 
enumerated above. For the sake of brevity, we will
not write down explicitly these alternative systems, especially since
in most cases they can be obtained by making a few changes in the
systems~(\ref{sysc}) for $\mb{R}\,.$
Finally, it is important to remark that the initial-data $\mb{{}_oU}$
in each case must satisfy, apart from the relations~(\ref{conf}),
the constraints.  Taking this into account, the
characterizations of the different equilibrium points that we are 
going to give will consist of the minimum conditions that, added to
the constraints, imply the relations defining such configurations.

\subsection{FLRW dust models}\label{flrw}

The FLRW models~\cite{FLRW} are the standard cosmological models.  They 
are motivated by the so-called {\em Cosmological Principle} in the sense 
that they are homogeneous and isotropic cosmological models (they have a 
six-dimensional group of motions).  In the case of dust, the line element 
can be expressed in terms of elementary functions as follows (see, 
e.g.,~\cite{HAEL}):
\[ \mbox{ds}^2 = - d\tau^2 + \frac{R^2(\tau)}{\left(1+\frac{k}{4}r^{2}
\right)^2} \delta_{\mu\nu}dy^\mu dy^\nu \hspace{3mm}
(r^2 \equiv \delta^{}_{\mu\nu}y^\mu y^\nu) \,, \]
where
\[ \left\{ \begin{array}{lclcl} R = (E/3)(\cosh t - 1)\,, & \hspace{3mm} 
& \tau = (E/3)(\sinh t - t)\,, & \hspace{3mm} & \mbox{if $k < 0$}\,, \\
R = t^2\,, & \hspace{3mm} & \tau = \textstyle{1\over3}t^3\,, &
\hspace{3mm} & \mbox{if $k = 0$}\,, \\
R = (-E/3)(1-\cos t)\,, &\hspace{3mm} & \tau = (-E/3)(t-\sin t)\,,
& \hspace{3mm} & \mbox{if $k > 0$}\,, \end{array}  \right. \]
$E$ is an arbitrary constant such that $\mbox{sign}(E)=-\mbox{sign}(k)$,
and the hypersurfaces $\Sigma(\tau)$ have constant curvature
(${}^3R=6kR^{-2}$).  There are several covariant spacetime 
characterizations of the FLRW models, and in particular of the dust 
subcase (see, e.g.,~\cite{KRAS}).  Here, we will be concerned with the 
following two characterizations:

(i) The {\em kinematical characterization}. In terms of the kinematical
quantities a dust FLRW model is characterized simply by the vanishing
of the shear
\[ \sigma_{ab}=0 \,. \]

(ii) The {\em Weyl tensor characterization}.  The FLRW dust models can be 
characterized by the vanishing of the Weyl tensor, or equivalently, by the 
vanishing of the gravito-electric and -magnetic fields
\[ E_{ab}=H_{ab}=0 \,.\]

From these spacetime characterizations we can see that the FLRW 
dust models constitute an equilibrium configuration in the dynamics of 
the IDMs,  finding at the same time an initial data characterization for
these models.  To that end, we need to consider the following relations 
\[ {\cal R}^1{}_{ab}\equiv\sigma_{ab}=0\,,~~
{\cal R}^2{}_{ab}\equiv E_{ab}=0\,,~~
{\cal R}^3{}_{ab}\equiv H_{ab}=0 \,. \]
Their evolution follows directly from~(\ref{base}-\ref{hwey})
\[ \dot{\cal R}^1{}_{\alpha\beta}=-\textstyle{2\over3}\Theta
{\cal R}^1{}_{\alpha\beta}-\sigma_{<\alpha}{}^\delta
{\cal R}^1{}_{\beta>\delta}-{\cal R}^2{}_{\alpha\beta} \,, \]
\[ \dot{\cal R}^2{}_{\alpha\beta}-\mbox{curl}({\cal R}^3)_{\alpha\beta} =
-\Theta{\cal R}^2{}_{\alpha\beta}+3\sigma_{<\alpha}{}^\delta
{\cal R}^2{}_{\beta>\delta}-\textstyle{1\over2}\rho
{\cal R}^1{}_{\alpha\beta}\,, \]
\[ \dot{\cal R}^3{}_{\alpha\beta}+\mbox{curl}({\cal R}^2)_{\alpha\beta} =
-\Theta{\cal R}^3{}_{\alpha\beta}+3\sigma_{<\alpha}{}^\delta
{\cal R}^3{}_{\beta>\delta}\,. \]
As is clear, $\mb{R}=\mb{0}$ is a solution, and since the equations for
$(\mb{U},\mb{R})$ constitute a FOSHS, it is the only one.  To sum up, 
we can say that {\em if the initial state belongs to the configuration 
characterized by the vanishing of the shear and the gravito-electric
field, the system will remain in this configuration.  Moreover, the
resulting spacetime belongs to the FLRW class}.  In this statement we 
have not taken into account the vanishing of the gravito-magnetic field 
because, as we have said before, we assume that the constraints hold.  
Apart from the vanishing of $H_{ab}$, the constraints imply 
\begin{equation}
 D_a\Theta = 0 \,, \hspace{3mm} D_a \rho = 0 \,, \label{flcc}
\end{equation}
that is, $\Theta$ and $\rho$ must be constant on the hypersurfaces
$\Sigma(\tau)$.

We can find another initial-data characterization of the FLRW models, 
equivalent to that given above, in terms of geometrical quantities only.  
It is based on the equations~(\ref{sffs}) and~(\ref{li7},\ref{li8}),
and the relations that characterize it are 
\[ {\cal R}^1{}_{ab}\equiv K_{ab}-\textstyle{1\over3}h_{ab}K = 0\,,
\hspace{3mm} {\cal R}^2{}_{ab}\equiv {}^3 S_{ab}=0 \,, \hspace{3mm} 
{\cal R}^3{}_{ab}\equiv \mbox{curl}(K)_{ab}= 0 \,. \]
As before, ${\cal R}^1{}_{ab}$, ${\cal R}^2{}_{ab}$ and the constraints
imply ${\cal R}^3{}_{ab}$.  The remaining conditions imposed by the
constraints, which are equivalent to (\ref{flcc}), are
\[ K|_{\Sigma(\tau)} = \mbox{constant} \,, \hspace{3mm} 
  {}^3R|_{\Sigma(\tau)} = \mbox{constant}\,.\]
All these relations together, tell us that the initial hypersurface
$\Sigma(\tau_o)$ must be of constant curvature and the second
fundamental form (which describes how it is immersed in the spacetime
manifold) must be proportional to the first fundamental form, with
the proportionality factor the trace $K$ of $K_{ab}\,,$ which is a constant.

\subsection{Spatially homogeneous IDMs}\label{spho}

This well-known class of spacetimes is characterized by the 
existence of a three-dimensional group of motions ($G_3$) simply
transitive on the hypersurfaces orthogonal to the fluid velocity, 
$\{\Sigma(\tau)\}$.  These models were studied systematically 
in~\cite{ELMC}, where a classification was given.  Following this 
work we can see that a spacetime characterization of these models 
is the following: ``a perfect-fluid cosmological model contains a 
three-dimensional group of motions simply transitive on three-surfaces 
orthogonal to the fluid velocity $\mb{u}$ if and only if there exits an
orthonormal basis $\{\mb{e}_0=\mb{u},\mb{e}_\alpha\}$ such that the
commutators satisfy
\[ \gamma^n_{pq} = \gamma^n_{pq}(\tau)
\hspace{5mm} \Longleftrightarrow \hspace{5mm}
\mb{\partial}_\alpha \gamma^n_{pq} = 0\,\mbox{.''} \]
The triad $\{\mb{e}_\alpha\}$ spans the surfaces of transitivity of
the group.

Starting from the above characterization we will prove the
following statement: {\em If the initial state is such that the 
quantities ${}_o\gamma^\alpha_{\beta\delta}$, ${}_o\Theta$, and
${}_o\sigma_{\alpha\beta}$ are constant, the system will stay in
this configuration.  The cosmological models constructed from the
development of initial data satisfying these conditions will correspond 
to spatially homogeneous (Bianchi) dust models}.
To show this, let us consider the following relations
\begin{equation} 
{\cal R}^1{}^{\lambda}{}_{\alpha\beta\delta}\equiv\mb{\partial}_\alpha 
\gamma^\lambda_{\beta\delta} = 0\,, \hspace{5mm} {\cal R}^2{}_\alpha 
\equiv  \mb{\partial}_\alpha \Theta = 0 \,, \hspace{5mm}
{\cal R}^3{}_{\alpha\beta\delta} \equiv \mb{\partial}_\alpha
\sigma_{\beta\delta} = 0 \,, \label{bia1}
\end{equation}
\begin{equation} 
 {\cal R}^4{}_{\alpha} \equiv
\mb{\partial}_\alpha \rho = 0  \,, \hspace{5mm} 
{\cal R}^5{}_{\alpha\beta\delta} \equiv \mb{\partial}_\alpha
E_{\beta\delta} = 0 \,, \hspace{5mm} {\cal R}^6{}_{\alpha\beta\delta} 
\equiv \mb{\partial}_\alpha H_{\beta\delta} = 0 \,. \label{bia2}
\end{equation}
Whereas the relations in~(\ref{bia1}) will implement the characterization,
the relations in~(\ref{bia2}) are auxiliary conditions needed for the
proof, but they can be derived from~(\ref{bia1}) and the constraints
(\ref{li4}-\ref{li8}).  
Using the evolution and constraint equations and the commutators of
spatial derivatives~(\ref{spco}) we obtain the following system of
evolution equations for $\mb{R}$
\begin{eqnarray}
\fl \dot{{\cal R}}^1{}^{\lambda}{}_{\alpha\beta\delta} =
-(\textstyle{1\over3}\Theta\delta_\alpha{}^\epsilon+
\sigma_\alpha{}^\epsilon){\cal R}^1{}^{\lambda}{}_{\epsilon\beta\delta}
+(\textstyle{1\over3}\Theta\delta^\epsilon{}_{[\beta}+
\sigma^\epsilon{}_{[\beta}){\cal R}^1{}^{\lambda}{}_{|\alpha|\delta]
\epsilon}+\gamma^\lambda_{\epsilon[\beta}(\textstyle{1\over3}
\delta^\epsilon{}_{\delta]}{\cal R}^2{}_\alpha+
{\cal R}^3{}_{|\alpha|\delta]}{}^\epsilon)
\nonumber \\
+2\delta^{\lambda\kappa}\delta_{\eta\gamma}\gamma^\gamma_{\kappa\epsilon}
\delta^{(\epsilon}{}_{[\beta}{\cal R}^3{}_{|\alpha|\delta]}{}^{\eta)}+
2\delta^{\lambda\kappa}\delta_{\eta\gamma}\delta^{(\epsilon}{}_{[\beta}
\delta^{\eta)}{}_{\delta]}{\cal R}^1{}^\gamma{}_{\alpha\kappa\epsilon}
-\varepsilon_{\beta\delta\epsilon}
{\cal R}^6{}_{\alpha}{}^{\lambda\epsilon}\,, \nonumber
\end{eqnarray}
\[\fl \dot{{\cal R}}^2{}_\alpha = -(\Theta\delta_\alpha{}^\epsilon+
\sigma_\alpha{}^\epsilon){\cal R}^2{}_\epsilon-
2\sigma^{\beta\delta}{\cal R}^3{}_{\alpha\beta\delta}-
\textstyle{1\over2}{\cal R}^4{}_\alpha \,, \]
\[\fl \dot{{\cal R}}^3{}_{\alpha\beta\delta}= -\textstyle{2\over3}
\sigma_{\beta\delta}{\cal R}^2{}_\alpha -(\Theta\delta_\alpha{}^\epsilon+
\sigma_\alpha{}^\epsilon){\cal R}^3{}_{\epsilon\beta\delta}-
2\sigma_{\langle\beta}{}^\epsilon
{\cal R}^3{}_{|\alpha|\delta\rangle\epsilon}-
{\cal R}^5{}_{\alpha\beta\delta}\,, \]
\[ \fl \dot{{\cal R}}^4{}_\alpha = -\rho{\cal R}^2{}_\alpha-
(\textstyle{4\over3}\Theta\delta_\alpha{}^\epsilon+
\sigma_\alpha{}^\epsilon){\cal R}^4{}_\epsilon \,, \]
\begin{eqnarray}
\fl \dot{{\cal R}}^5{}_{\alpha\beta\delta}=-E_{\beta\delta}{\cal R}^2{}_\alpha
-\textstyle{1\over2}\rho{\cal R}^3{}_{\alpha\beta\delta}+
3E_{\langle\beta}{}^\epsilon{\cal R}^3{}_{|\alpha|\delta\rangle\epsilon}-
\textstyle{1\over2}\sigma_{\beta\delta}{\cal R}^4{}_\alpha-
(\textstyle{4\over3}\Theta\delta_\alpha{}^\epsilon+
\sigma_\alpha{}^\epsilon){\cal R}^5{}_{\epsilon\beta\delta} 
\nonumber \\ +3\sigma_{\langle\beta}{}^\epsilon
{\cal R}^5{}_{|\alpha|\delta\rangle\epsilon}
+\gamma^\epsilon_{\alpha\lambda}
\varepsilon^{\lambda\kappa}{}_{\langle\beta}
{\cal R}^6{}_{|\epsilon|\delta\rangle\kappa}+
\varepsilon_{\lambda\kappa\langle\beta}\mb{\partial}^\lambda
{\cal R}^6{}_{|\alpha|\delta\rangle}{}^{\kappa}\nonumber \,, 
\end{eqnarray}
\begin{eqnarray}
\fl \dot{{\cal R}}^6{}_{\alpha\beta\delta}=-H_{\beta\delta}
{\cal R}^2{}_\alpha+3H_{\langle\beta}{}^\epsilon
{\cal R}^3{}_{|\alpha|\delta\rangle\epsilon}-(\textstyle{4\over3}\Theta
\delta_\alpha{}^\epsilon+\sigma_\alpha{}^\epsilon)
{\cal R}^6{}_{\epsilon\beta\delta}+3\sigma_{\langle\beta}{}^\epsilon
{\cal R}^6{}_{|\alpha|\delta\rangle\epsilon}\nonumber \\
-\gamma^\epsilon_{\alpha\lambda}\varepsilon^{\lambda\kappa}
{}_{\langle\beta}{\cal R}^5{}_{|\epsilon|\delta\rangle\kappa}-
\varepsilon_{\lambda\kappa\langle\beta}\mb{\partial}^\lambda
{\cal R}^5{}_{|\alpha|\delta\rangle}{}^{\kappa}\nonumber \,. 
\end{eqnarray}
As we can see, only the two last equations contain spatial derivatives,
and their structure is essentially the same as that of
equations (\ref{ewey},\ref{hwey}). Then the extended
system for $(\mb{U},\mb{R})$ is a FOSHS, with $\mb{R}=\mb{0}$ the only
solution.  This shows that Bianchi
models constitute an equilibrium configuration characterized 
by~(\ref{bia1}).

This result provides an initial-data characterization of the dust 
Bianchi models within the class of IDMs, but it says nothing about the 
particular classes of Bianchi models (see~\cite{ELMC} for a
classification).  We can study which kind of equilibrium configurations
particular Bianchi models may constitute by refining the characterization
given in expressions~(\ref{bia1},\ref{bia2}).  To that end, we will have
to add more conditions depending on the class of Bianchi models under
consideration.  As an example, let us consider the Bianchi I dust 
models~\cite{BIAI}, in which the $G_3$ group is Abelian and the 
line-element is given by
\begin{equation}  
\mbox{ds}^2 = -d\tau^2 + \sum_{\alpha=1}^3\tau^{2p_\alpha}
(\tau-{}_o\tau)^{2(2/3-p_\alpha)}(dy^\alpha)^2 \,, \label{dsbi}
\end{equation}
where ${}_o\tau$, $p_\alpha$ are constants such that $\sum_{\alpha=1}^3 
p^{}_\alpha = \sum_{\alpha=1}^3 p^2_\alpha = 1$.
These models can be characterized by adding the following conditions:
\[ {\cal R}^7{}_\alpha\equiv a_\alpha = 0 \,, \hspace{5mm}
{\cal R}^8{}_{\alpha\beta}\equiv n_{\alpha\beta} = 0 \,, \hspace{5mm}
{\cal R}^9{}_{\alpha\beta}\equiv H_{\alpha\beta} = 0 \,. \]
Here ${\cal R}^9{}_{\alpha\beta}$ is a consequence of
${\cal R}^3{}_{\alpha\beta\delta}$ [Eq.~(\ref{bia1})],
${\cal R}^7{}_{\alpha\beta}$, ${\cal R}^8{}_{\alpha\beta}$ and
the constraint~(\ref{li4}).  We can obtain the following evolution
equations for these quantities:
\[ \fl \dot{\cal R}^7{}_\alpha = -\textstyle{1\over3}\Theta
{\cal R}^7{}_\alpha+\textstyle{1\over2}\sigma_\alpha{}^\beta
{\cal R}^7{}_\beta-\textstyle{1\over2}\varepsilon_{\alpha}
{}^{\beta\delta}\sigma_\beta{}^\lambda{\cal R}^8{}_{\delta\lambda}\,, \]
\begin{eqnarray*} 
\fl \dot{\cal R}^8{}_{\alpha\beta}=-\textstyle{1\over3}\Theta
{\cal R}^8{}_{\alpha\beta}-\sigma_{<\alpha}{}^\lambda
{\cal R}^8{}_{\beta>\lambda}+\delta_{\alpha\beta}\sigma^{\lambda\delta}
{\cal R}^8{}_{\lambda\delta}+\textstyle{1\over2}\sigma_{\alpha\beta}
{\cal R}^8{}^\lambda{}_\lambda+\sigma_\lambda{}^{<\alpha}
\varepsilon^{\beta>\lambda\delta}{\cal R}^7{}_\delta-
{\cal R}^9{}_{\alpha\beta}\,, 
\end{eqnarray*}
\begin{eqnarray*} 
\fl \dot{\cal R}^9{}_{\alpha\beta}=-\Theta{\cal R}^9{}_{\alpha\beta}
+3\sigma_{<\alpha}{}^\lambda{\cal R}^9{}_{\beta>\lambda}-
\varepsilon^{\lambda\delta}{}_{<\alpha}
{\cal R}^5{}_{|\lambda|\beta>\delta} \,. 
\end{eqnarray*}
The whole system of evolution equations for the quantities $\mb{U}$
and ${\cal R}^I$ ($I=1,\ldots,9$) still constitutes a FOSHS, and
${\cal R}^I=0$ is a solution, the only possible one.
Therefore, we can say that {\em conditions~(\ref{bia1}) and
$({\cal R}^7{}_{\alpha\beta},{\cal R}^8{}_{\alpha\beta})$ determine
an equilibrium configuration in the dynamics of the IDMs, which
corresponds to the Bianchi I dust models}.  In a similar way and
using the information given in~\cite{ELMC} we can study which
equilibrium configurations come from the other classes of Bianchi 
models.

The characterizations given above do not have a covariant translation.
However, in some cases it is possible to get a covariant characterization,
an example being Bianchi I IDMs.  As in previous cases we start
from covariant spacetime characterizations.  For Bianchi I models two such
characterizations follow from~\cite{SOPL}.  The first one is determined by
the vanishing of the spatial covariant derivative of the shear
\[ D_a \sigma_{bc} = 0 \,. \]
The second one says that an IDM belongs to the class of the
Bianchi I dust models if and only if the hypersurfaces $\Sigma(\tau)$ 
are flat 
\[ {}^3R_{ab} = 0 \,, \]
and $E_{ab}$ (or equivalently $\sigma_{ab}$) is non-degenerate
(algebraically general).  Taking this into account, a covariant
initial-data characterization of the Bianchi I dust models is given by
\begin{equation} 
{\cal R}^1{}_{abc}\equiv D_aK_{bc}=D_a(\sigma_{bc}+\textstyle{1\over3}
\Theta h_{bc})=0\,, \hspace{4mm} {\cal R}^2{}_{ab}\equiv {}^3 R_{ab}=0\,.
\label{sccb}
\end{equation}
The evolution equations for these relations form a closed system:
\begin{eqnarray} 
\dot{\cal R}^1{}_{abc}=-K{\cal R}^1{}_{abc}-K_{bc}{\cal R}^1{}_{ad}{}^d
-K_a{}^d{\cal R}^1{}_{dbc}-2K_{(b}{}^d{\cal R}^1{}_{c)ad}+
2K_{d(b}{\cal R}^{1d}{}_{c)a} \nonumber \\
-D_a{\cal R}^2{}_{bc}+\textstyle{1\over4}h_{bc}\left(D_a{\cal R}^2{}_d{}^d
+2K{\cal R}^1{}_{ad}{}^d-2K^{ed}{\cal R}^1{}_{aed}\right) \,, \nonumber
\end{eqnarray}
\begin{eqnarray} 
\dot{\cal R}^2{}_{ab}= -K{\cal R}^2{}_{ab}+K_{(a}{}^c{\cal R}^2{}_{b)c}
-\textstyle{1\over2}K_{ab}{\cal R}^2{}_c{}^c+\mbox{curl}[
\varepsilon_{cd(a}{\cal R}^{1c}{}_{b)}{}^d] \nonumber \\
+h_{ab}\left(\textstyle{1\over6}K{\cal R}^2{}_c{}^c
-K^{cd}{\cal R}^2{}_{cd}\right)\,, \nonumber
\end{eqnarray}
where $K_{ab}$ must be understood as given in terms of the expansion
and shear~(\ref{sffs}).  Again, $\mb{R}=\mb{0}$ is a solution, but now
the system of evolution equations we have for $(\mb{U},\mb{R})$ is not
a FOSHS.  However, using the procedure described above we can show that
$\mb{R}=\mb{0}$ is the only possible solution, showing that~(\ref{sccb})
constitutes a stationary configuration associated with the Bianchi I 
models~(\ref{dsbi}).

\subsection{Szekeres dust models}\label{szmo}
The Szekeres dust cosmological models~\cite{SZEK} are inhomogeneous
models. As was shown in~\cite{BOST}, they do
not have in general any Killing vector field. From this point of
view they can be considered to be generic cosmological models, but from 
other points of view, as for instance the algebraic classification of 
spacetimes, they are not so generic since they are Petrov type D models,
a special algebraic case.
The line element can be written in the following diagonal form
\begin{equation}
\fl \mbox{ds}^2 = -d\tau^{2} + \Sigma^{-2}U(\tau,y^1)(dy^1)^2+
A^2(\tau,y^\alpha)\left[(dy^2)^2+ k^2(y^2,y^3)(dy^3)^2\right] \,,
\label{dssz}
\end{equation}
where $\Sigma$ coincides with the repeated shear eigenvalue
$\sigma^2=\sigma^3$.  From their spacetime characterization
(see~\cite{KRAS} and references therein)
\begin{equation} 
[\sigma,E]_a=0 \,, \hspace{3mm} H_{ab}= 0 \,, \hspace{3mm} 
E_{ab}~(\mbox{or}~\sigma_{ab})~\mbox{degenerated}\,, \label{stcs}
\end{equation}
we can see that the evolution equations~(\ref{base}-\ref{hwey}) become
local equations~\cite{SILE}, that is, the spatial derivatives disappear
(there are no fields propagating in these spacetimes).  Hence,
the evolution equations are ordinary differential equations.

We can obtain an initial-data characterization of these models starting
from the spacetime characterization~(\ref{stcs}).
Taking into account the form of the third condition in~(\ref{stcs}),
the use of an orthonormal basis adapted to the fluid velocity
simplifies considerably the relations we will use to get the
characterization.  These relations are
\begin{equation} 
{\cal R}^1{}_{-,1,2,3}\equiv(\sigma_-,\sigma_1,\sigma_2,\sigma_3)=0 
\,, ~
{\cal R}^2{}^{}_{11,22,33,23}\equiv(n_{11},n_{22},n_{33},n_{23})=0 
\,, \label{shnn}
\end{equation}
\begin{equation} 
{\cal R}^3{}_{-,1,2,3}\equiv(E_-,E_1,E_2,E_3)=0 \,, \label{trel} 
\end{equation}
\begin{equation} 
{\cal R}^4{}_{\alpha\beta}\equiv H_{\alpha\beta}=0 ~~ 
\Longleftrightarrow ~~ {\cal R}^4{}_{+,-,1,2,3}\equiv
(H_+,H_-,H_1,H_2,H_3)=0 \,, \label{ere4}
\end{equation}
\begin{equation} 
 {\cal R}^5{}_{\alpha\beta}\equiv M_{\alpha\beta}=0 ~~ 
\Longleftrightarrow ~~ {\cal R}^5{}_{+,-,1,2,3}\equiv
(M_+,M_-,M_1,M_2,M_3)=0 \,, 
\end{equation}
\begin{equation} 
{\cal R}^6{}_{\alpha\beta}\equiv N_{\alpha\beta}=0 ~~ 
\Longleftrightarrow ~~ {\cal R}^6{}_{+,-,1,2,3}\equiv
(N_+,N_-,N_1,N_2,N_3)=0 \,, \label{defn}
\end{equation}
where we have introduced the following definitions $M_{ab}\equiv
\mbox{curl}(E)_{ab}\,,$ and $N_{ab}\equiv \mbox{curl}(H)_{ab}$, which
become new constraints
\begin{equation}
{\cal C}^7{}_{ab} \equiv \mbox{curl}(E)_{ab}-M_{ab} = 0 \,, \label{con7}
\end{equation}
\[ {\cal C}^8{}_{ab} \equiv \mbox{curl}(H)_{ab}-N_{ab} = 0 \,. \]
For these quantities we did not give evolution equations, but we can
find them using the evolution equations~(\ref{base}-\ref{hwey}).
The result can be expressed as follows:
\[\fl \dot{M}_{ab}=-\textstyle{4\over3}\Theta M_{ab} +\textstyle{3\over2}
\sigma_{<a}{}^c[\sigma,H]_{b>c} - \textstyle{1\over2}\rho H_{ab} +
3E_{<a}{}^cH_{b>c}+\mbox{curl}(N)_{ab} + {\cal F}^1{}_{ab} \,, \]
\[\fl \dot{N}_{ab}=-\textstyle{4\over3}\Theta N_{ab}+3H_{<a}{}^cH_{b>c}-
\mbox{curl}(M)_{ab}+ {\cal F}^2{}_{ab} \,, \]
where 
\begin{equation}
\fl {\cal F}^1{}_{ab}\equiv\textstyle{3\over2}
\varepsilon_{cd<a}E_{b>}{}^c D_e\sigma^{de}+\textstyle{3\over2}
\varepsilon_{cd<a}\sigma_{b>}{}^cD_eE^{de}+3\mbox{curl}\,
(\sigma\cdot E)_{ab}-\sigma_e{}^c\varepsilon_{cd<a}D^eE_{b>}{}^d
+{\cal CC}_{ab} \,, \label{puf1}
\end{equation}
\[\fl {\cal F}^2{}_{ab}\equiv\textstyle{3\over2}
\varepsilon_{cd<a}H_{b>}{}^c D_e\sigma^{de}+3\mbox{curl}\,
(\sigma\cdot H)_{ab}-\sigma_e{}^c\varepsilon_{cd<a}D^eH_{b>}{}^d
+{\cal CC}'{}_{ab} \,,\]
and 
\begin{equation} 
(\sigma\cdot E)_{ab}\equiv\sigma_{<a}{}^cE_{b>c}\,,~~~~~
(\sigma\cdot H)_{ab}\equiv\sigma_{<a}{}^cH_{b>c}\,. \label{tepr}
\end{equation}
Moreover, the quantities ${\cal CC}_{ab}$ and ${\cal CC}'{}_{ab}$ represent 
combinations of the constraints, which vanish when the constraints 
hold. Here, we have used the well-known fact that adding combinations
of the constraints to the evolution equations does not change the
dynamics, but they can be useful for technical reasons and can 
modify the type of differential equations that we obtain. We will use
this fact to find the initial-data characterization of the Szekeres 
models.  Now we are ready to compute the evolution equations for
the quantities ${\cal R}^I$.  They can be expressed in the following 
form
\[ \fl \dot{\cal R}^1{}_-= -\textstyle{2\over3}(\Theta+\sigma_+)
{\cal R}^1{}_- -\textstyle{1\over{2\sqrt{3}}}(\sigma_2{\cal R}^1{}_2+
\sigma_3{\cal R}^1{}_3)-{\cal R}^3{}_- \,,\]
\[ \fl \dot{\cal R}^1{}_1=-\textstyle{2\over3}(\Theta+\sigma_+){\cal R}^1{}_1
-\textstyle{1\over\sqrt{3}}\sigma_2{\cal R}^1{}_3-{\cal R}^3{}_1 \,,\]
\[ \fl \dot{\cal R}^1{}_2=-\textstyle{1\over3}(2\Theta+\sigma_+ + 
\sqrt{3}\sigma_-){\cal R}^1{}_2-\textstyle{1\over\sqrt{3}}\sigma_1
{\cal R}^1{}_3-{\cal R}^3{}_2 \,, \]
\[ \fl \dot{\cal R}^1{}_3=-\textstyle{1\over3}(2\Theta-\sigma_+ + 
\sqrt{3}\sigma_-){\cal R}^1{}_3-\textstyle{1\over\sqrt{3}}\sigma_1
{\cal R}^1{}_2-{\cal R}^3{}_3 \,, \]
\begin{eqnarray}
\fl \dot{\cal R}^2{}_{11}=-\textstyle{1\over3}(\Theta+\sigma_+)
{\cal R}^2{}_{11}+\textstyle{1\over\sqrt{3}}\sigma_-
({\cal R}^2{}_{22}-{\cal R}^2{}_{33}) \nonumber \\
\fl ~~~
+\textstyle{1\over\sqrt{3}}(n_{12}{\cal R}^1{}_3+n_{13}{\cal R}^1{}_2+
2n_{23}{\cal R}^1{}_1)
+\textstyle{1\over\sqrt{3}}(a_3{\cal R}^1{}_3-a_2{\cal R}^1{}_2)+
\textstyle{2\over3}{\cal R}^4{}_+\,, \label{en11}
\end{eqnarray}
\begin{eqnarray}
\fl \dot{\cal R}^2{}_{22}=\textstyle{1\over6}(2\Theta-\sigma_+-\sqrt{3}
\sigma_-){\cal R}^2{}_{22}-
\textstyle{1\over{2\sqrt{3}}}(\sqrt{3}\sigma_+-\sigma_-)
({\cal R}^2{}_{11}-{\cal R}^2{}_{33})\nonumber \\ 
\fl ~~~
+\textstyle{1\over\sqrt{3}}(n_{12}{\cal R}^1{}_3+n_{23}{\cal R}^1{}_1+
2n_{13}{\cal R}^1{}_2)+\textstyle{1\over\sqrt{3}}(a_1{\cal R}^1{}_1-
a_3{\cal R}^1{}_3)-\textstyle{1\over3}({\cal R}^4{}_++\sqrt{3}
{\cal R}^4{}_-)\,, \label{en22}
\end{eqnarray}
\begin{eqnarray}
\fl \dot{\cal R}^2{}_{33}=-\textstyle{1\over6}(2\Theta-\sigma_++
\sqrt{3}\sigma_-){\cal R}^2{}_{33}-\textstyle{1\over{2\sqrt{3}}}
(\sqrt{3}\sigma_++\sigma_-)({\cal R}^2{}_{11}-{\cal R}^2{}_{22})
\nonumber \\  
\fl ~~~
+\textstyle{1\over\sqrt{3}}(n_{13}{\cal R}^1{}_2+
n_{23}{\cal R}^1{}_1+2n_{12}{\cal R}^1{}_3)
+\textstyle{1\over\sqrt{3}}(a_2{\cal R}^1{}_2-a_1{\cal R}^1{}_1)
-\textstyle{1\over3}({\cal R}^4{}_+-\sqrt{3}{\cal R}^4{}_-)\,, 
\label{en33}
\end{eqnarray}
\begin{eqnarray}
\fl \dot{\cal R}^2{}_{23}=-\textstyle{1\over3}(\Theta+\sigma_+)
{\cal R}^2{}_{23}-\textstyle{1\over{2\sqrt{3}}}(n_{13}{\cal R}^1{}_3
+n_{12}{\cal R}^1{}_2-n_{11}{\cal R}^1{}_1) \nonumber \\ 
-\textstyle{1\over{2\sqrt{3}}}(2a_1{\cal R}^1{}_--a_2{\cal R}^1{}_3+
a_3{\cal R}^1{}_2)-\textstyle{1\over\sqrt{3}}{\cal R}^4{}_1\,, 
\label{en23}
\end{eqnarray}
\[\fl \dot{\cal R}^3{}_-=-(\Theta-\sigma_+){\cal R}^3{}_- -\textstyle
{\sqrt{3}\over2}(\sigma_2{\cal R}^3{}_2-\sigma_3{\cal R}^3{}_3) 
+(E_+-\textstyle{1\over2}\rho){\cal R}^1{}_- +{\cal R}^6{}_-\,,\]
\[\fl \dot{\cal R}^3{}_1=-(\Theta-\sigma_+){\cal R}^3{}_1+\textstyle
{\sqrt{3}\over2}(\sigma_2{\cal R}^3{}_3+\sigma_3{\cal R}^3{}_2)
+(E_+-\textstyle{1\over2}\rho){\cal R}^1{}_1 +{\cal R}^6{}_1\,,\]
\begin{eqnarray*}
\fl \dot{\cal R}^3{}_2=-[\Theta+\textstyle{1\over2}(\sigma_++
\sqrt{3}\sigma_-)]{\cal R}^3{}_2+\textstyle{\sqrt{3}\over2}
(\sigma_3{\cal R}^3{}_1+\sigma_1{\cal R}^3{}_3)  \\
\fl ~~~ -\textstyle{1\over2}
(E_++\sqrt{3}E_-+\rho){\cal R}^1{}_2+{\cal R}^6{}_2\,, 
\end{eqnarray*}
\begin{eqnarray*}
\fl \dot{\cal R}^3{}_3=-[\Theta+\textstyle{1\over2}(\sigma_+-
\sqrt{3}\sigma_-)]{\cal R}^3{}_3+\textstyle{\sqrt{3}\over2}
(\sigma_2{\cal R}^3{}_1+\sigma_1{\cal R}^3{}_2) \\
\fl ~~~ -\textstyle{1\over2}
(E_+-\sqrt{3}E_-+\rho){\cal R}^1{}_3+{\cal R}^6{}_3\,, 
\end{eqnarray*}
\[\fl \dot{\cal R}^4{}_{\alpha\beta}=-\Theta{\cal R}^4{}_{\alpha\beta}+
3\sigma_{<\alpha}{}^\delta
{\cal R}^4{}_{\beta>\delta}-{\cal R}^5{}_{\alpha\beta}\,, \]
\begin{eqnarray}
\fl \dot{\cal R}^5{}_{\alpha\beta}= 
-\textstyle{4\over3}\Theta{\cal R}^5{}_{\alpha\beta}+\textstyle{3\over2}
\sigma_{<\alpha}{}^\delta[\sigma,{\cal R}^4]_{\beta>\delta}
-\textstyle{1\over2}\rho {\cal R}^4{}_{\alpha\beta} +
3E_{<\alpha}{}^\delta{\cal R}^4{}_{\beta>\delta}+
\mbox{curl}({\cal R}^6){}_{\alpha\beta} + 
{\cal F}^1{}_{\alpha\beta} \,, \nonumber
\end{eqnarray}
\begin{eqnarray}
\fl \dot{\cal R}^6{}_{\alpha\beta}=
-\textstyle{4\over3}\Theta{\cal R}^6{}_{\alpha\beta}+
3H_{<\alpha}{}^\delta{\cal R}^4{}_{\beta>\delta}-\textstyle{3\over2}
\mbox{div}(\sigma)^\delta\varepsilon_{\delta\gamma<\alpha}
{\cal R}^4{}_{\beta>}{}^\gamma
+3\mbox{curl}(\sigma\cdot{\cal R}^4)_{\alpha\beta}\nonumber \\ 
-\sigma_\gamma{}^\delta\varepsilon_{\delta\kappa<\alpha}
D^\gamma{\cal R}^4{}_{\beta>}{}^\kappa-\mbox{curl}
({\cal R}^5){}_{\alpha\beta} \,. \nonumber
\end{eqnarray}
The main point is to show that $\mb{R}=\mb{0}$ is a solution for this 
system of equations.  As we can see, this question depends only on the 
form of the different terms that make up the tensor
${\cal F}^1{}_{\alpha\beta}$ [see Eq.~(\ref{puf1})].  The explicit
expressions for its components are given in~\ref{appa}, where the
combination of constraints ${\cal CC}_{ab}$ has been chosen in the
following way: ${\cal CC}_+ = {\cal CC}_- = {\cal CC}_1 = 0$, and
[see Eqs.~(\ref{li4}) and~(\ref{con7})]
\begin{equation}
{\cal CC}_2 = \textstyle{3\over2}E_+{\cal C}^2{}_2+\textstyle{11\over2}
\sigma_+{\cal C}^7{}_2 \,, \label{ecc2}
\end{equation}
\begin{equation}
{\cal CC}_3 = \textstyle{3\over2}E_+{\cal C}^2{}_3+\textstyle{11\over2}
\sigma_+{\cal C}^7{}_3 \,. \label{ecc3}
\end{equation}
With this choice we have shown in~\ref{appa} that
${\cal F}^1{}_{\alpha\beta}=0$ when ${\cal R}^I=0$.   In the case of the
evolution equation for ${\cal R}^6{}_{\alpha\beta}$ we have not added any
constraint, i.e.,  we have chosen ${\cal CC}'{}_{\alpha\beta}=0$.
To sum up, we can find a system such that ${\cal R}^I=0$ is a solution.
The whole system of evolution equations for $(\mb{U},\mb{R})$ is not
a FOSHS system, but following the discussion above in this section
we can modify the system for $\mb{R}$ to show that ${\cal R}^I=0$ is 
the only possible solution.
The conclusion is given in the following statement: {\em If the initial
state of the matter fluid belongs to the configuration defined by the 
following covariant conditions} 
\begin{equation} 
[\sigma,E]_a=0 \,, \hspace{3mm} H_{ab}=0 \,, \hspace{3mm} 
E_{ab}~(\mbox{{\em or}}~\sigma_{ab})~\mbox{{\em degenerate}}\,,
\hspace{3mm}  \mbox{curl}(E)_{ab} = 0 \,, \label{casz}
\end{equation}
{\em then it will belong to the configuration for all later time}. It
is worth noting that conditions (\ref{shnn}-\ref{defn}) have a
covariant translation.
Apart from defining
an equilibrium configuration, conditions~(\ref{casz}) provide an
initial-data characterization of the Szekeres models (the Szafron
models~\cite{SZAF} with constant pressure in the case of a non-zero
cosmological constant).  The remaining conditions are auxiliary, since
they follow from (\ref{casz}) and the constraints,  or they can be obtained
just by fixing the remaining freedom in the choice of the triad
$\{\mb{e}_\alpha\}\,.$

On the other hand, we can have a covariant characterization equivalent
to~(\ref{casz}) expressed only in terms of geometrical quantities.
It is given by the following relations
\begin{equation} 
[K,{}^3R]_a=0 \,, \hspace{5mm} \mbox{curl}(K)_{ab} = 0 \,, 
\hspace{5mm} C_{ab} = 0 \,, \label{gcas}
\end{equation}
where $C_{ab}$ is proportional to the conformally-invariant 
Cotton-York curvature tensor~\cite{YORK} associated with the hypersurfaces
$\Sigma(\tau)$, $\tilde{\beta}^{ab}$. The
proportionality factor is given by
\[ \tilde{\beta}^{ab} = h^{5/3}C^{ab}\,, \hspace{4mm} h\equiv
[\mbox{det}(h_{ab})]^{1/2} \,. \]
One can check that $C_{ab}$ can be expressed in terms of the 
trace-free part of the Ricci tensor of the hypersurfaces 
$\Sigma(\tau)$ as follows
\[ C_{ab} = \mbox{curl}({}^3S)_{ab} \,.\]
$C_{ab}$ has essentially the same properties as the Cotton-York tensor:
i) It is spatial, symmetric and trace-free. ii) It is transverse
(divergence-free): $D_aC^{ab}=0$.
iii) It vanishes if and only if the hypersurfaces $\Sigma(\tau)$ are
locally conformally-flat.  The vanishing of this tensor has been
sometimes related with the absence of gravitational waves~\cite{BEEO}.
For the Szekeres models this interpretation seems to be correct
since $C_{ab}$ vanishes, and as we have said before, there is
an absence of propagation signals in these models (the evolution 
equations are completely local).

The first two conditions in~(\ref{gcas}) are clearly a direct
translation of the first two conditions in (\ref{casz}).  Therefore,
to show this equivalence we only have to show that the third conditions 
are equivalent.  To that end,  we can use the constraint~(\ref{li8}) to 
write $C_{ab}$ in the following form
\begin{eqnarray}
C_{ab}  & = & \mbox{curl}(E)_{ab}-
\textstyle{1\over3}\Theta H_{ab} +\sigma^{d}{}_e\varepsilon_{cd<a}D^e
\sigma_{b>}{}^c+\textstyle{3\over2}\varepsilon_{cd<a}\sigma_{b>}{}^c
D_e\sigma^{de} \nonumber \\
& \equiv & \mbox{curl}(E)_{ab}-\textstyle{1\over3}\Theta H_{ab}+
{\cal T}_{ab} \,. \label{deft}
\end{eqnarray}
Then the problem reduces to checking that the terms in ${\cal T}_{ab}$
vanish when the relations (\ref{shnn}-\ref{defn}) hold.   From~(\ref{deft}),
${\cal T}_{ab}$ contains only the shear and its first derivatives
and hence we only have to check that its components do not contain
any term of the type $\sigma_+\mb{\partial_\alpha}\sigma_+$ or
$\Gamma\,\sigma^2_+\,,$ where $\Gamma=a_\alpha\,,$ $n_{12}\,,$ or
$n_{13}\,.$  Since the terms contained in ${\cal T}_{ab}$ have the same
structure as those in ${\cal F}^1{}_{ab}$, we can use the calculations done
for ${\cal F}^1{}_{ab}$ (see \ref{appa}).  For the components
${\cal T}_+\,,$ ${\cal T}_-\,,$ and ${\cal T}_1\,,$ it is a matter of
algebra to check that the terms of the type mentioned above do not appear,
and for the components ${\cal T}_2$ and ${\cal T}_3$ we need to use the
components $2$ and $3$ of the constraint ${\cal C}^2{}_{ab}$
[Eq.~(\ref{li4})] to see that they cancel.
These calculations finish the proof that the conditions (\ref{gcas})
provide an initial-data characterization of the Szekeres models equivalent
to that given in~(\ref{casz}).

\subsection{IDMs with flat hypersurfaces}\label{plam}

The spacetime characterization of these models is obvious; they are
defined by the vanishing of the Ricci tensor of the hypersurfaces 
$\Sigma(\tau)$, i.e.,
\begin{equation} 
{}^3R_{ab} = 0 \,. \label{vacu}
\end{equation}
From a geometrical point of view these models can be considered as
inhomogeneous generalizations of the standard Einstein-de Sitter
cosmological model~\cite{EIDS}.  Their relevance is enhanced by recent
observations~\cite{BOOM}, which suggest that the spatial geometry of the
local universe is close to flat.  For the case of a vanishing cosmological
constant ($\Lambda=0$), the IDMs satisfying~(\ref{vacu}) were determined
in~\cite{SOPL}.  There are two classes of solutions, both having a
vanishing gravito-magnetic field, $H_{ab}=0$.  The first one consists
of the Bianchi I dust models~(\ref{dsbi}), which are Petrov type I models.
The second class, which has a degenerate gravito-electric tensor 
(Petrov type D), is the subclass of the Szekeres models~\cite{SZEK}
satisfying~(\ref{vacu}), which were given in~\cite{BOSP}.  Since we
have already studied the first class,  we will focus here on the 
second class, which is composed of two families of solutions.
The line element of the first family is given by [$(y^\alpha)=(x,y,z)$]
\begin{equation} 
\mbox{ds}^2 = -d\tau^2 + \left[ (1+A y+B z)\tau+C\right]^2
\tau^{-\textstyle{2\over3}}dx^2 + \tau^{\textstyle{4\over3}}
(dy^2+dz^2) \,, \label{cla1}
\end{equation}
where $A$, $B$ and $C$ are arbitrary functions of $x$. The line
element of the second family is
\begin{equation}
\fl \mbox{ds}^2 = -d\tau^2 + \frac{V^2[\partial_x(\ln U)]^2}
{(\tau-{}_o\tau)^{\textstyle{2\over3}}}\left[\tau-{}_o\tau + \frac{2}{3}
\frac{\partial_x({}_o\tau)}{\partial_x(\ln U)}\right]^2 dx^2 
+ U^2(\tau-{}_o\tau)^{\textstyle{4\over3}}
(dy^2+dz^2) \,, \label{cla2}
\end{equation}
where 
\[ U = V W\,, \hspace{5mm} 
W=\left\{ a\left[y^2+z^2\right]+2b y+2c z+d\right\}^{-1} \,,\]
and $(a,b,c,d)$, $V$ and ${}_o\tau$ are any functions of $x$ such
that $ad-b^2-c^2=1$.
Remarkably, despite the particular character of these two solutions,
both share the property of having no Killing vectors fields in general,
like the generic Szekeres metric~(\ref{dssz}).

Now we will show that the IDMs~(\ref{cla1},\ref{cla2}) constitute an
equilibrium configuration in the dynamics of IDMs, and we will also find 
its characterization.  To that end, we need to consider the following 
relations: (\ref{shnn}),~(\ref{ere4}) and the alternative to~(\ref{trel}),
\[ \fl {\cal R}^3{}_{ab}\equiv {}^3R_{ab}=0 ~~ \Longleftrightarrow ~~
{\cal R}^3{}^a{}_a\equiv {\cal R}^3={}^3R=0 ~~ \mbox{and} ~~
{\cal R}^3{}_{<ab>}\equiv {\cal S}^3{}_{ab}={}^3S_{ab}=0 \,.\]
The evolution equations for them are
given by the equations~(\ref{en11}-\ref{en23}), which do not change, and 
by the equations
\[ \dot{\cal R}^1{}_{-,1,2,3} = -\Theta{\cal R}^1{}_{-,1,2,3} 
- {\cal S}^3{}_{-,1,2,3} \,, \]
\[ \dot{\cal R}^3 = -\textstyle{2\over3}\Theta{\cal R}^3 -
2\sigma^{\alpha\beta}{\cal S}^3{}_{\alpha\beta} \,, \] 
\[ \dot{\cal S}^3{}_{\alpha\beta} = -\textstyle{2\over3}\Theta
{\cal S}^3{}_{\alpha\beta}+\sigma_{<\alpha}{}^\delta
{\cal S}^3{}_{\beta>\delta}-\textstyle{1\over6}\sigma_{\alpha\beta}
{\cal R}^3+\mbox{curl}({\cal R}^4){}_{\alpha\beta}\,, \] 
\[ \dot{\cal R}^4{}_{\alpha\beta} =  -\textstyle{4\over3}\Theta
{\cal R}^4{}_{\alpha\beta}+3\sigma_{<\alpha}{}^\delta
{\cal R}^4{}_{\beta>\delta}-\mbox{curl}({\cal S}^3){}_{\alpha\beta}+
{\cal T}_{\alpha\beta}\,, \] 
where ${\cal T}_{\alpha\beta}$ is defined by Eq.~(\ref{deft}).
Therefore, we have a closed system for which ${\cal R}^I=0$
is a solution, which can be shown to be unique. Then we can
say that {\em if initially the spatial geometry is flat, the
gravito-magnetic field vanishes and the shear is degenerate, the
system will stay in this configuration}.  Again, we have been able to
extract a covariant initial-data characterization from the relations
that define the equilibrium configuration.

\section{Remarks and discussion}\label{sec5}

In this paper we have considered the general dynamics of IDMs
from an Initial-Value Problem perspective.  The equations
governing these models constitute an infinite-dimensional dynamical
system due to the presence of spatial gradients.   We have seen that
a variety of well-known models, from the homogeneous and isotropic
FLRW spacetimes to the inhomogeneous Szekeres cosmological models,
constitute equilibrium configurations of the dynamics (invariant sets).
It remains to find the characterizations of some important IDMs, as
the remaining Bianchi classes of IDMs and the Lema\^{\i}tre-Tolman-Bondi
dust models (the initial-value problem approach made in~\cite{SUTR}
could lead to such characterization).
Moreover, we have found the relations characterizing these configurations,
which in most cases can be expressed in a covariant way.  At the 
same time, these relations provide an initial-data characterization
of the IDMs considered.  The information obtained together with the
study of the evolution of the different physical quantities clarifies 
what the dynamical content of these IDMs is, as well as the role
of the different variables and the spatial gradients.

This work must be considered as a first step towards the implementation
of a more general programme to understand the full dynamics of the 
gravitational instability mechanism.  In that respect, two interesting
issues to be elucidated are the question of how a triaxial configuration
will evolve in collapse (among the models treated in this paper, only
homogeneous models have an algebraically general shear and gravito-electric
field), and the role of non-local effects due to the appearance of the
curls of the gravito-electric and -magnetic fields.
Such a programme would entail a method to prescribe generic
initial data.  This means finding solutions
of the constraints for any possible initial distribution of matter.
The programme would also include a numerical code to evolve this initial data,
i.e. to construct its development, and capable of extracting the
physical information.  The development of these two points, which
does not seem an easy task, could be complemented by the information
coming from studies based on perturbative methods (perturbations of a
background or iterative approaches such as the long wave-length approximation
scheme).

The results of this paper provide useful information for the development
of such a programme.  First, they can help us in the prescription of 
initial data: we know the role of the variables and
what kind of initial data we have to avoid in order to study new
behaviours.   Moreover, since the characterizations given are solutions to
the constraints, this information can be helpful in solving them 
for other different cases.  On the other hand, we can use the
characterizations to check numerical codes and to identify 
attractors, repellers and asymptotic states of the evolution.

Finally, it is worth noting that the present work
can be extended to models with a more general energy-momentum
content.  In this sense, a straightforward generalization
would be the study of irrotational perfect-fluid models,
including in this way the effects of the pressure and hence,
the inclusion of sound propagation.  Furthermore, it would also
be interesting to consider fluids with rotation to study their
effects in the gravitational instability mechanism.  This would 
require introducing a spacelike foliation of the spacetime
not related with the fluid velocity, since in that case
it does not generate orthogonal hypersurfaces.

\ack

I wish to thank Professor R. Maartens for a careful reading of the
manuscript and M.\'A.G. Bonilla and Professors D. Kramer and
R. Maartens for interesting discussions.  I wish also to thank
the Alexander von Humboldt Foundation for financial support and the
Institute for Theoretical Physics of the Jena University for hospitality
during the first stages of this work.  Currently, the author is supported
by the European Commission (contract HPMF-CT-1999-00149).

%
%

\appendix

\section{Explicit expressions for the components of 
${\cal F}^1{}_{ab}$\label{appa}}

To complete the proof of the characterization of the Szekeres 
cosmological models in terms of initial data given in subsection
\ref{szmo}, we need to show that the quantity ${\cal F}^1{}_{ab}$
[see Eq.~(\ref{puf1})] vanishes when ${\cal R}^I=0\,.$
To that end we have to work out the components of 
${\cal F}^1{}_{\alpha\beta}$, the projection of ${\cal F}^1{}_{ab}$
with respect to the orthonormal triad $\{\mb{e}_\alpha\}$.
The terms ${\cal CC}_{ab}$ and ${\cal CC}'{}_{ab}$, denoting
combinations of the constraints, have been chosen as follows:
${\cal CC}'{}_{ab}=0\,,$ ${\cal CC}_+ = {\cal CC}_- = {\cal CC}_1 = 0$,
and ${\cal CC}_2$ and ${\cal CC}_3$ are given in
Eqs.~(\ref{ecc2},\ref{ecc3}).

In order to simplify the large expressions for the components
of ${\cal F}^1{}_{ab}$ we will use the following notation:
For any components $A_+$ and $A_-$ of a tensor $A_{\alpha\beta}$
\[ \hat{A}_\pm \equiv \sqrt{3}\,A_+ \pm A_- \,,~~~~
   \tilde{A}_\pm \equiv (A_+ \pm \sqrt{3}\,A_-)/\sqrt{3} \,. \]
The components of the tensor $(\sigma\cdot E)_{ab}\equiv{\cal P}_{ab}$ [see
Eq.~(\ref{tepr})] can be written as
\[ \fl {\cal P}_+ = -\textstyle{1\over3}(\sigma_+E_+-\sigma_-{\cal R}^3{}_-)-
\textstyle{1\over6}(\sigma_2{\cal R}^3{}_2+\sigma_3{\cal R}^3{}_3-
2\sigma_1{\cal R}^3{}_1) \equiv  -\textstyle{1\over3}\sigma_+E_+
+ {\cal Q}_+\,, \]
\[ \fl {\cal P}_- = \textstyle{1\over3}(E_+{\cal R}^1{}_-+\sigma_+
{\cal R}^3{}_-)-\textstyle{1\over{2\sqrt{3}}}(\sigma_2{\cal R}^3{}_2-
\sigma_3{\cal R}^3{}_3) \,, \]
\[ \fl {\cal P}_1 = \textstyle{1\over3}(\sigma_+{\cal R}^3{}_1+E_+
{\cal R}^1{}_1)+\textstyle{1\over{2\sqrt{3}}}(\sigma_3{\cal R}^3{}_2+
\sigma_2{\cal R}^3{}_3) \,, \]
\[ \fl {\cal P}_2 = -\textstyle{1\over{2\sqrt{3}}}(\tilde{\sigma}_+
{\cal R}^3{}_2+\tilde{E}_+{\cal R}^1{}_2-\sigma_3{\cal R}^3{}_1-
\sigma_1{\cal R}^3{}_3) \,, \]
\[ \fl {\cal P}_3 = -\textstyle{1\over{2\sqrt{3}}}(\tilde{\sigma}_-
{\cal R}^3{}_3+\tilde{E}_-{\cal R}^1{}_3-\sigma_2{\cal R}^3{}_1-
\sigma_1{\cal R}^3{}_2) \,. \]
As we can see, the quantities ${\cal Q}_+\,,$ ${\cal P}_-\,,$
${\cal P}_1\,,$  ${\cal P}_2\,,$ and ${\cal P}_3$ vanish when
${\cal R}^1{}_I={\cal R}^3{}_I=0\,.$

Taking all these definitions into account, the components of 
${\cal F}^1{}_{ab}$ can be written as follows
\begin{eqnarray}
\fl {\cal F}^1{}_+ & = & \textstyle{3\over4}\left\{ 
[ (\mb{\partial_2}-3a_2)\tilde{\sigma}_+ +n_{13}\hat{\sigma}_-
+(n_{11}-n_{33})\sigma_2+n_{12}\sigma_1-n_{23}\sigma_3]{\cal R}^3{}_2- 
\right. \nonumber \\
\fl & & [ (\mb{\partial_3}-3a_3)\tilde{\sigma}_- -n_{12}\hat{\sigma}_+
-(n_{11}-n_{22})\sigma_3-n_{13}\sigma_1+n_{23}\sigma_2]{\cal R}^3{}_3+
\nonumber \\
\fl & & \left. 2[\mb{\partial_1}\sigma_{[3}+\mb{\partial_{[3}}\sigma_{|1|}
-3(\sigma_1a_{[3}+a_1\sigma_{[3})]{\cal R}^3{}_{2]} + (\mb{\sigma}
\leftrightarrow \mb{E}\,, {\cal R}^3 \leftrightarrow {\cal R}^1)\right\} +
\nonumber \\
\fl & & \textstyle{{3\sqrt{3}}\over2}\left\{(\mb{\partial_3}-a_3+n_{12})
{\cal P}_3-(\mb{\partial_2}-a_2-n_{13}){\cal P}_2-2n_{23}{\cal P}_1-
3{\cal P}_+{\cal R}^2{}_{11}- \right. \nonumber \\
\fl & & \left. {\cal P}_-({\cal R}^2{}_{22}-{\cal R}^2{}_{33})
\right\}+ \nonumber \\
\fl & & \textstyle{1\over2}\left\{ \tilde{\sigma}_+[\mb{\partial_2}
{\cal R}^3{}_2+(a_1+n_{23}){\cal R}^3{}_1+(a_3-n_{12}){\cal R}^3{}_3+
\textstyle{1\over2}\hat{E}_-({\cal R}^2{}_{11}-{\cal R}^2{}_{22}+
{\cal R}^2{}_{33})] - \right. \nonumber \\
\fl & & \tilde{\sigma}_-[\mb{\partial_3}{\cal R}^3{}_3+(a_1-n_{23})
{\cal R}^3{}_1+(a_2+n_{13}){\cal R}^3{}_2-\textstyle{1\over2}\hat{E}_+
({\cal R}^2{}_{11}+{\cal R}^2{}_{22}-{\cal R}^2{}_{33})]- \nonumber \\
\fl & & [2(\mb{\partial_{[2}}E_{3]}+\sqrt{3}\,n_{23}E_++a_1E_-)+
(a_2+n_{13})E_3-(a_3-n_{12})E_2+n_{11}E_1]{\cal R}^1{}_1 - \nonumber \\
\fl & & [\mb{\partial_1}E_3-(a_2-n_{13})\hat{E}_+-(a_3+n_{12})E_1+
\textstyle{1\over2}(n_{11}-n_{22}-n_{33})E_2]{\cal R}^1{}_2+ \nonumber \\
\fl & & \left. [\mb{\partial_1}E_2-(a_3+n_{12})\hat{E}_--(a_2-n_{13})E_1-
\textstyle{1\over2}(n_{11}-n_{22}-n_{33})E_3]{\cal R}^1{}_3 \right\} 
\nonumber \,,  \label{fmas}
\end{eqnarray}
 
\begin{eqnarray}
\fl {\cal F}^1{}_- & = & -\textstyle{\sqrt{3}\over4}\left\{ 
[ (\mb{\partial_2}-3a_2)\tilde{\sigma}_+ +n_{13}\hat{\sigma}_-
+(n_{11}-n_{33})\sigma_2+n_{12}\sigma_1-n_{23}\sigma_3]{\cal R}^3{}_2+ 
\right. \nonumber \\ 
\fl & & [ (\mb{\partial_3}-3a_3)\tilde{\sigma}_- -n_{12}\hat{\sigma}_+
-(n_{11}-n_{22})\sigma_3-n_{13}\sigma_1+n_{23}\sigma_2]{\cal R}^3{}_3- 
\nonumber \\
\fl & & 4[(\mb{\partial_{(2}}-3a_{(2})\sigma_{3)}
-\textstyle{1\over\sqrt{3}}(\mb{\partial_1}
-3a_1)\sigma_++n_{23}\sigma_--\textstyle{1\over2}(n_{22}-n_{33})\sigma_1+
\nonumber \\
\fl & &  +\textstyle{1\over2}(n_{13}\sigma_3-n_{12}\sigma_2)]
{\cal R}^3{}_1+
2[(\mb{\partial_1}-3a_1)\sigma_{(2}+(\mb{\partial_{(2}}-3a_{(2})\sigma_{|1|}]
{\cal R}^3{}_{3)} + \nonumber \\
\fl & & \left. (\mb{\sigma}
\leftrightarrow \mb{E}\,, {\cal R}^3 \leftrightarrow {\cal R}^1)\right\} -
\nonumber \\
\fl & & \textstyle{3\over2}\left\{(\mb{\partial_3}-a_3-3n_{12}){\cal P}_3+
(\mb{\partial_2}-a_2+3n_{13}){\cal P}_2-2(\mb{\partial_1}-a_1){\cal P}_1-
\right. \nonumber \\
\fl & & \left. \sqrt{3}{\cal P}_+({\cal R}^2{}_{22}-{\cal R}^2{}_{33})+
{\cal P}_-({\cal R}^2{}_{11}-2{\cal R}^2{}_{22}-2{\cal R}^2{}_{33})
\right\}- \nonumber \\
\fl & & \textstyle{1\over{2\sqrt{3}}}\left\{\textstyle{4\over\sqrt{3}} 
\sigma_+[\mb{\partial_1}{\cal R}^3{}_1+(a_2-n_{13}){\cal R}^3{}_2+
(a_3+n_{12}){\cal R}^3{}_3-(n_{11}-n_{22}-n_{33}){\cal R}^3{}_-]+
\right. \nonumber \\
\fl & & \tilde{\sigma}_+[\mb{\partial_2}{\cal R}^3{}_2+(a_1+n_{23})
{\cal R}^3{}_1+(a_3-n_{12}){\cal R}^3{}_3+
\textstyle{1\over2}\hat{E}_-({\cal R}^2{}_{11}-{\cal R}^2{}_{22}+
{\cal R}^2{}_{33})] + \nonumber \\
\fl & & \tilde{\sigma}_-[\mb{\partial_3}{\cal R}^3{}_3+(a_1-n_{23})
{\cal R}^3{}_1+(a_2+n_{13}){\cal R}^3{}_2-\textstyle{1\over2}\hat{E}_+
({\cal R}^2{}_{11}+{\cal R}^2{}_{22}-{\cal R}^2{}_{33})]+ \nonumber \\
\fl & & [2(\mb{\partial_{(2}}E_{3)}+\sqrt{3}\,a_1E_++n_{23}E_-)-
(n_{22}-n_{33})E_1-(a_3-n_{12})E_2- \nonumber \\
\fl & & (a_2+n_{13})E_3]{\cal R}^1{}_1 +
[(\mb{\partial_1}+2a_1-2n_{23})E_3-(2\mb{\partial_3}+a_3+n_{12})E_1
- \nonumber \\
\fl & & (a_2-n_{13})\hat{E}_++4(a_2+n_{13})E_- -
\textstyle{1\over2}(n_{11}+3n_{22}-n_{33})E_2]{\cal R}^1{}_2+
\nonumber \\
\fl & & [(\mb{\partial_1}+2a_1+2n_{23})E_2-(2\mb{\partial_2}+a_2-n_{13})E_1-
(a_3+n_{12})\hat{E}_--4(a_3-n_{12})E_- +\nonumber \\
\fl & & \left. \textstyle{1\over2}(n_{11}-n_{22}+3n_{33})E_3]{\cal R}^1{}_3 
\right\}  \,, \label{fmenos}
\end{eqnarray}

\begin{eqnarray}
\fl {\cal F}^1{}_1 & = & \textstyle{\sqrt{3}\over4}\left\{ 
[ (\mb{\partial_2}-3a_2)\tilde{\sigma}_+ +n_{13}\hat{\sigma}_-
+(n_{11}-n_{33})\sigma_2+n_{12}\sigma_1-n_{23}\sigma_3]{\cal R}^3{}_3- 
\right. \nonumber \\ 
\fl & & [ (\mb{\partial_3}-3a_3)\tilde{\sigma}_- -n_{12}\hat{\sigma}_+
-(n_{11}-n_{22})\sigma_3-n_{13}\sigma_1+n_{23}\sigma_2]{\cal R}^3{}_2- 
\nonumber \\
\fl & & 4[(\mb{\partial_{(2}}-3a_{(2})\sigma_{3)}-\textstyle{1\over\sqrt{3}}
(\mb{\partial_1}-3a_1)\sigma_+ +n_{23}\sigma_- 
-\textstyle{1\over2}(n_{22}-n_{33})\sigma_1+\nonumber \\
\fl & & \textstyle{1\over2}(n_{13}\sigma_3-n_{12}\sigma_2)]{\cal R}^3{}_-
-2[(\mb{\partial_1}-3a_1)\sigma_{[2}+(\mb{\partial_{[2}}-3a_{[2})\sigma_{|1|}]
{\cal R}^3{}_{3]} + \nonumber \\
\fl & & \left. (\mb{\sigma}\leftrightarrow \mb{E}\,, 
{\cal R}^3 \leftrightarrow {\cal R}^1)\right\} +
\nonumber \\
\fl & & \textstyle{3\over2}\left\{(\mb{\partial_3}-a_3-3n_{12}){\cal P}_2-
(\mb{\partial_2}-a_2+3n_{13}){\cal P}_3+2(\mb{\partial_1}-a_1){\cal P}_- -
\right. \nonumber \\
\fl & & \left. 2\sqrt{3}{\cal P}_+{\cal R}^2{}_{23}+(n_{11}-2n_{22}-2n_{33})
{\cal P}_1\right\}+ \nonumber \\
\fl & & \textstyle{1\over{2\sqrt{3}}}\left\{
\textstyle{4\over\sqrt{3}}\sigma_+[\mb{\partial_1}{\cal R}^3{}_- 
+(a_2-n_{13}){\cal R}^3{}_3-(a_3+n_{12}){\cal R}^3{}_2+(n_{11}-n_{22}-n_{33})
{\cal R}^3{}_1]+ \right.  \nonumber \\
\fl & & \tilde{\sigma}_+[\mb{\partial_2}{\cal R}^3{}_3+(a_1+n_{23})
\hat{\cal R}^3{}_+ -(a_3-n_{12}){\cal R}^3{}_2+
\textstyle{1\over2}(n_{11}-n_{22}+n_{33}){\cal R}^3{}_1] - 
\nonumber \\
\fl & & \tilde{\sigma}_-[\mb{\partial_3}{\cal R}^3{}_2+(a_1-n_{23})
\hat{\cal R}^3{}_- -(a_2+n_{13}){\cal R}^3{}_3-\textstyle{1\over2}
(n_{11}+n_{22}-n_{33}){\cal R}^3{}_1]- 
\nonumber \\
\fl & & [(\mb{\partial_2}-a_2-n_{13})E_2-(\mb{\partial_3}-a_3+n_{12})E_3
+\sqrt{3}n_{11}E_+ +(n_{22}-n_{33})E_-+ \nonumber \\
\fl & & 2n_{23}E_1]{\cal R}^1{}_1 + [(\mb{\partial_1}-2a_1+2n_{23})E_2-
2\mb{\partial_3}E_- +(a_3+n_{12})\hat{E}_- - \nonumber \\
\fl & & (3a_2+5n_{13})E_1 - \textstyle{1\over2}(n_{11}+3n_{22}-
n_{33})E_3]{\cal R}^1{}_2+ \nonumber \\
\fl & & [(\mb{\partial_1}+2a_1+2n_{23})E_3 - 2\mb{\partial_2}E_- -
(a_2-n_{13})\hat{E}_+ +(3a_3-5n_{12})E_1 - \nonumber \\
\fl & & \left. \textstyle{1\over2}(n_{11}-n_{22}+3n_{33})E_2]{\cal R}^1{}_3 
\right\}  \,,   \label{funo}
\end{eqnarray}

\begin{eqnarray}
\fl {\cal F}^1{}_2 & = & \textstyle{\sqrt{3}\over4}\left\{
[ 2(\mb{\partial_{(1}}-3a_{(1})\sigma_{2)}+ (\mb{\partial_3}-3a_3)
\tilde{\sigma}_- -n_{12}\hat{\sigma}_+ -(n_{11}-n_{22})\sigma_3+
n_{23}\sigma_2- \right. \nonumber \\
\fl & & n_{13}\sigma_1]{\cal R}^3{}_1-2[ (\mb{\partial_{(2}}-
3a_{(2})\sigma_{3)}-
\textstyle{1\over\sqrt{3}}(\mb{\partial_1}-3a_1)\sigma_+
+n_{23}\sigma_- -\textstyle{1\over2}(n_{22}-n_{33})\sigma_1+ \nonumber \\
\fl & & \textstyle{1\over2}(n_{13}\sigma_3-n_{12}\sigma_2)]{\cal R}^3{}_3-
\hat{E}_-[2(\mb{\partial_{(1}}-3a_{(1}){\cal R}^1{}_{3)}
+(\mb{\partial_2}-3a_2-n_{13}){\cal R}^1{}_- + \nonumber \\
\fl & & (n_{11}-n_{33}){\cal R}^1{}_2+n_{12}{\cal R}^1{}_1-n_{23}
{\cal R}^1{}_3]-\sqrt{3}E_+[(\mb{\partial_2}-a_2-3n_{13}){\cal R}^1{}_- +
\nonumber \\
\fl & & (\mb{\partial_3}-a_3+3n_{12}){\cal R}^1{}_1- (\mb{\partial_1}-a_1-
3n_{23}){\cal R}^1{}_3- (n_{22}-2n_{11}-2n_{33}){\cal R}^1{}_2]+
\nonumber \\
\fl & & \left.
\textstyle{1\over\sqrt{3}}[(\mb{\partial_2}-3a_2+3n_{13})\sigma_+]
{\cal R}^3{}_- +(\mb{\sigma}\leftrightarrow \mb{E}\,,
{\cal R}^3 \leftrightarrow {\cal R}^1)\right\} + \nonumber \\
\fl & & \textstyle{3\over2}\left\{(\mb{\partial_2}-a_2-3n_{13})(
\sqrt{3}{\cal Q}_+-{\cal P}_-) - (\mb{\partial_3}-a_3+3n_{12})
{\cal P}_1+(\mb{\partial_1}-a_1-3n_{23}){\cal P}_3 -
\right. \nonumber \\
\fl & & \left. (n_{22}-2n_{11}-2n_{33}){\cal P}_2\right\}-
\textstyle{1\over6}\left\{3(\partial_2E_+){\cal R}^1{}_- +
9E_+{\cal R}^4{}_2+11\sigma_+{\cal R}^5{}_2\right\}+ \nonumber \\
\fl & & \textstyle{1\over{2\sqrt{3}}}\left\{
\tilde{\sigma}_-[\mb{\partial_3}{\cal R}^3{}_1 -2(a_2+n_{13})
\hat{\cal R}^3{}_- -(a_1-n_{23}){\cal R}^3{}_3+
\textstyle{1\over2}(n_{11}+n_{22}-n_{33}){\cal R}^3{}_2] + \right.
\nonumber \\
\fl & & \tilde{\sigma}_+[\mb{\partial_2}{\cal R}^3{}_-
-2(a_3-n_{12})\hat{\cal R}^3{}_1 +2(a_1+n_{23}){\cal R}^3{}_3+
2(n_{11}-n_{22}+n_{33}){\cal R}^3{}_2]+
\nonumber \\
\fl & & \textstyle{1\over\sqrt{3}}\sigma_+[ (3\mb{\partial_1}-
a_1-3n_{23}){\cal R}^3{}_3-(\mb{\partial_2}+a_2-5n_{13})
{\cal R}^3{}_- -(\mb{\partial_3}+a_3+5n_{12}){\cal R}^3{}_1-
\nonumber \\
\fl & & (n_{11}+3n_{33}){\cal R}^3{}_2] -
[\mb{\partial_3}\hat{E}_- -(\mb{\partial_2}+2a_2+2n_{13})E_1
-2(a_3-n_{12})E_- - \nonumber \\
\fl & & (3a_1-5n_{23})E_2 +\textstyle{1\over2}(3n_{11}+n_{22}-
n_{33})E_3]{\cal R}^1{}_1+ \nonumber \\
\fl & & [(\mb{\partial_1}+a_1-n_{23})E_1 - (\mb{\partial_3}-a_3-n_{12})
E_3 +2n_{13}E_2 + (n_{11}-n_{22}-n_{33})E_- - \nonumber \\
\fl & & \textstyle{1\over2}(n_{11}+n_{22}-n_{33})\hat{E}_+]{\cal R}^1{}_2
- [\mb{\partial_1}\hat{E}_- -(\mb{\partial_2}-2a_2+2n_{13})E_3
+(a_1+n_{23})\hat{E}_+ + \nonumber \\
\fl & & \left. (3a_3+5n_{12})E_2 -\textstyle{1\over2}(n_{11}-n_{22}-
3n_{33})E_1]{\cal R}^1{}_3\right\}  \,, \label{fdos}
\end{eqnarray}

\begin{eqnarray}
\fl {\cal F}^1{}_3 & = & -\textstyle{\sqrt{3}\over4}\left\{
[ 2(\mb{\partial_{(1}}-3a_{(1})\sigma_{3)}+ (\mb{\partial_2}-3a_2)
\tilde{\sigma}_+ +n_{13}\hat{\sigma}_- +(n_{11}-n_{33})\sigma_2+
n_{12}\sigma_1- \right. \nonumber \\
\fl & & n_{23}\sigma_3]{\cal R}^3{}_1-2[ (\mb{\partial_{(2}}-3a_{(2})
\sigma_{3)}-\textstyle{1\over\sqrt{3}}(\mb{\partial_1}-3a_1)\sigma_+
+n_{23}\sigma_- -\textstyle{1\over2}(n_{22}-n_{33})\sigma_1
+ \nonumber \\
\fl & & \textstyle{1\over2}(n_{13}\sigma_3-n_{12}\sigma_2)]{\cal R}^3{}_2-
\hat{E}_-[2(\mb{\partial_{(1}}-3a_{(1}){\cal R}^1{}_{2)}
-(\mb{\partial_3}-3a_3+n_{12}){\cal R}^1{}_- -  \nonumber \\
\fl & & (n_{11}-n_{22}){\cal R}^1{}_3+n_{23}{\cal R}^1{}_2-n_{13}
{\cal R}^1{}_1]+ \sqrt{3}E_+[(\mb{\partial_3}-a_3+3n_{12}){\cal R}^1{}_- -
\nonumber \\
\fl & & (\mb{\partial_2}-a_2-3n_{13}){\cal R}^1{}_1+ (\mb{\partial_1}-a_1+
3n_{23}){\cal R}^1{}_2-(n_{33}-2n_{11}-2n_{22}){\cal R}^1{}_3]-
\nonumber \\
\fl & & \left. \textstyle{1\over\sqrt{3}}[(\mb{\partial_3}-3a_3-
3n_{12})\sigma_+]{\cal R}^3{}_- +(\mb{\sigma}\leftrightarrow \mb{E}\,,
{\cal R}^3 \leftrightarrow {\cal R}^1)\right\} - \nonumber \\
\fl & & \textstyle{3\over2}\left\{(\mb{\partial_3}-a_3+3n_{12})(
\sqrt{3}{\cal Q}_++{\cal P}_-) - (\mb{\partial_2}-a_2-3n_{13})
{\cal P}_1+(\mb{\partial_1}-a_1+3n_{23}){\cal P}_2 -
\right. \nonumber \\
\fl & & \left. (n_{33}-2n_{11}-2n_{22}){\cal P}_3\right\}-
\textstyle{1\over6}\left\{3(\partial_3E_+){\cal R}^1{}_- +
9E_+{\cal R}^4{}_3+11\sigma_+{\cal R}^5{}_3\right\}- \nonumber \\
\fl & & \textstyle{1\over{2\sqrt{3}}}\left\{
\tilde{\sigma}_+[\mb{\partial_2}{\cal R}^3{}_1 +2(a_3-n_{12})
\hat{\cal R}^3{}_- -(a_1+n_{23}){\cal R}^3{}_2-
\textstyle{1\over2}(n_{11}-n_{22}+n_{33}){\cal R}^3{}_3] + \right.
\nonumber \\
\fl & & \tilde{\sigma}_-[\mb{\partial_3}{\cal R}^3{}_-
+2(a_2+n_{13})\hat{\cal R}^3{}_1 -2(a_1-n_{23}){\cal R}^3{}_2+
2(n_{11}+n_{22}-n_{33}){\cal R}^3{}_3]+
\nonumber \\
\fl & & \textstyle{1\over\sqrt{3}}\sigma_+[ (3\mb{\partial_1}-
a_1+3n_{23}){\cal R}^3{}_2+(\mb{\partial_3}+a_3+5n_{12})
{\cal R}^3{}_- -(\mb{\partial_2}+a_2-5n_{13}){\cal R}^3{}_1+
\nonumber \\
\fl & & (n_{11}+3n_{22}){\cal R}^3{}_3] -
[\mb{\partial_2}\hat{E}_+ +(\mb{\partial_3}-2a_3+2n_{12})E_1
+2(a_2+n_{13})E_- - \nonumber \\
\fl & & (3a_1+5n_{23})E_3 -\textstyle{1\over2}(3n_{11}-n_{22}+n_{33})
E_2]{\cal R}^1{}_1- \nonumber \\
\fl & & [\mb{\partial_1}\hat{E}_+ +(\mb{\partial_3}+2a_3+2n_{12})E_2
+(a_1-n_{23})\hat{E}_- +(3a_2-5n_{13})E_3 +\nonumber \\
\fl & & \textstyle{1\over2}(n_{11}-3n_{22}-n_{33})E_1]{\cal R}^1{}_2+
[(\mb{\partial_1}-a_1-n_{23})E_1 - (\mb{\partial_2}-a_2+n_{13})
E_2 + \nonumber \\
\fl & & \left. 2n_{12}E_3 - (n_{11}-n_{22}-n_{33})E_- -\textstyle{1\over2}
(n_{11}-n_{22}+n_{33})\hat{E}_-]{\cal R}^1{}_3 \right\}  \,. \label{ftres}
\end{eqnarray}
Expressions~(\ref{fmas}-\ref{ftres}) show explicitly that ${\cal F}^1{}_{ab}=0$
when ${\cal R}^I=0\,.$

%
%


\begin{thebibliography}{99}
\bibitem{ANAL} These definitions are analogous to those for vector fields:
$\mbox{div}(A)\equiv D_aA^a$ and $\mbox{curl}(A)_a\equiv\varepsilon_{abc}
D^bA^c$
\bibitem{STRH} A system of the form~(\ref{fosh}) is said to be {\em strongly}
hyperbolic if there is a 1-form $\mb{t}$ such that all the eigenvalues of the
matrix $t_a{\cal M}^a$ are real and it has a complete set of eigenvectors
\end{thebibliography}
\end{document}